\documentclass{article}

\usepackage[utf8]{inputenc}
\usepackage{amsmath}
\usepackage{fullpage}
\usepackage{relsize}
\usepackage{amssymb}
\usepackage{amsfonts}
\usepackage{graphicx}
\usepackage{tikz}
\usepackage{float}
\usepackage{authblk}
\usepackage[english]{babel}
\usepackage{mathtools}
\usepackage{array}

\usepackage{subfig}

\usepackage[a4paper, margin=0.75in]{geometry}

\usepackage[square,comma,numbers,sort&compress]{natbib}
\usepackage{hyperref}

\usetikzlibrary{arrows.meta}
\usetikzlibrary{calc}
\usetikzlibrary{shapes.misc}

\renewcommand{\i}{\mathrm{i}}
\newcommand{\e}{\mathrm{e}}
\newcommand{\eps}{\epsilon}

\renewcommand{\d}{\,\mathrm{d}}
\newcommand{\diff}[2]{\frac{\mathrm{d} #1}{\mathrm{d} #2}}

\usetikzlibrary{arrows.meta}
\usetikzlibrary{calc}
\usetikzlibrary{decorations.pathmorphing, patterns,shapes}
\usetikzlibrary{plotmarks}

\title{Stokes' phenomenon in continuous limits of discrete Painlev\'{e} I}
\author[1]{Christopher J. Lustri\footnote{Electronic address: christopher.lustri@sydney.edu.au}}
\author[2]{John R. King}
\affil[1]{School of Mathematics and Statistics, The University of Sydney, New South Wales 2050, Australia}
\date{}
\affil[2]{School of Mathematical Sciences, University of Nottingham, Nottingham, NG7 2RD, United Kingdom}
\date{}

\begin{document}
\maketitle

\abstract{We use exponential asymptotic analysis to identify the relevance of Stokes' phenomenon to integrability in discrete systems. We study Stokes' phenomenon in two discrete problems with the same (leading-order) continuous limit, a finite-difference discretisation of the first continuous Painlev\'{e} equation and the first discrete Painlev\'{e} equation, as well as a family of differential equation associated with each discrete problem. This analysis reveals two important observations. Firstly, the orderly behaviour that characterises Stokes' phenomenon in discrete equations emerges naturally from corresponding continuous differential equations as the order of the latter increases, although this is not apparent at low orders. Secondly, Stokes' phenomenon vanishes in the continuum limit of the integrable discrete equation, but not the non-integrable discrete equation. This means that subdominant exponentials do not appear in the integrable equation, and therefore do not cause moveable singularities to form in the solution. The results are clarified further by consideration of one-parameter family of difference equations that interpolates between the two considered in detail.}

\section{Introduction}

We consider two discrete variants of a scaled version of the first Painlev\'{e} equation, 
\begin{equation}\label{e:P1m}
\diff{^2y}{z^2} + 3y^2 = -2z.
\end{equation}
The first discrete equation is obtained by approximating the differential term using a second-order central derivative. The second equation is the first discrete Painlev\'{e} equation (dPI), which is integrable \cite{ablowitz2000extension,ramani1991discrete}. Both of these equations tend to \eqref{e:P1m} in the appropriate continuum limit \cite{cresswell1999discrete} when represented as an infinite-order differential equation{.}

This purpose of this study is to answer two questions:
\begin{enumerate}
\item How is Stokes' phenomenon seen in discrete equations consistent with that seen in continuous differential equations with finite order?
\item What are the differences in Stokes' phenomena between integrable and non-integrable discrete equations that have the same continuum limit?
\end{enumerate}
We will resolve the first question by studying Stokes' phenomenon in the infinite-order equation for the finite-difference discretisation of \eqref{e:P1m} and comparing it with the finite-order differential equations obtained by truncating the finite-difference expansions. We will resolve the second question by comparing Stokes' phenomenon from the finite-difference discretisation with the corresponding behaviour in the integrable first discrete Painlev\'{e} equation.

Recent studies have shown that discretisation of continuous systems causes asymptotic effects to appear in the solution that depend on the choice of discretisation. In \cite{joshi2019generalized}, the authors observed that discretising the KdV (Korteweg-de Vries) equation using a finite-difference scheme causes exponentially small ripples to appear around the classical soliton solution. The authors of \cite{moston2023nanoptera} studied this phenomenon in a discretised nonlinear Schr\"{o}dinger (NLS) equation. They identified the same behaviour, and found that the choice of discretisation, including properties such as the discretisation order and the relative step size in space and time, can cause the ripples to change in form or even vanish entirely. 

This asymptotic behaviour is a consequence of Stokes' phenomenon. Stokes' phenomenon describes behaviour seen in asymptotic solutions to singularly-perturbed equations, whereby terms that are exponentially small in the asymptotic limit appear and disappear as important curves in the complex plane, known as Stokes lines, are crossed \cite{Berry1991,Boyd1999}. The solutions also contain anti-Stokes lines, across which exponentially small terms grow to become non-negligible. 

To calculate the effects of Stokes' phenomenon, we utilise asymptotic techniques that can capture behaviour that is exponentially small in the asymptotic limit, known as exponential asymptotics. If a divergent asymptotic series, such as those typically obtained from singularly perturbed equations, is truncated optimally, the remainder term is generally exponentially small in the asymptotic limit \cite{Berry1991,Boyd1999}. This remainder term may be studied directly, and describes an exponentially small asymptotic contributions to the solution. This idea was introduced by Berry \cite{Berry1988,Berry} and Berry \& Howls \cite{BerryHowls1990}, who used such methods to determine the behaviour of special functions such as the Airy function. 

In this study, we will apply an exponential asymptotic method for studying discrete equations in their continuum limit developed by King \& Chapman \cite{king_chapman_2001}, which relies on writing the discrete equation as an infinite-order ordinary differential equation. This method built on the techniques developed by Olde Daalhuis et al. \cite{Daalhuis}  for linear differential equations, and by Chapman et al. \cite{Chapman} for nonlinear differential equations. We will use this method to locate Stokes lines in the two discrete equations and to calculate the exponentially small terms that appear as they are crossed. These methods were used in \cite{king_chapman_2001} to study dislocations in Frenkel-Kontorova chains, in \cite{joshi2015,joshi2017,joshi2019} to study the validity of asymptotic solutions to discrete Painlev\'{e} I and II, and $q$-Painlev\'{e} I, and in \cite{Alfimov,joshi2019generalized,moston2023nanoptera} to study exponentially small waves in discretised KdV and NLS equations. 

In these analyses, Stokes' phenomenon led to the appearance of an infinite number of {exponential terms} with exponents $-\chi/\epsilon$, where the values of $\chi'$ were evenly spaced along the imaginary axis, typically occurring at $\chi' = 2M\pi\i$ for $M \in \mathbb{Z}$. This behaviour is not seen in corresponding continuous equations \cite{joshi2019generalized,moston2023nanoptera}. In this study, we will clarify how the behaviour of {exponential terms} in discrete systems emerges from the associated infinite-dimensional differential equations. 

An important class of discrete equations is integrable nonlinear difference equations, such as the discrete Painlev\'{e} equations. Integrability in discrete equations is typically associated with conditions such as singularity confinement and algebraic entropy (see \cite{ablowitz2000extension} for a detailed discussion of indicators for discrete integrability). We will not concentrate on discrete integrability conditions, but rather the emergence of continuous markers of integrability when taking appropriate continuum limits; however, we will discuss singularity confinement as a marker of discrete integrability for two significant discrete equations in Appendix \ref{S.A2Delta}.

Joshi \& Lustri \cite{joshi2015} and Joshi, Luu \& Lustri \cite{joshi2017} used exponential asymptotic techniques to calculate the effect of Stokes' phenomenon in the far-field solutions to discrete Painlev\'{e} equations I and II. Other previous studies \cite{bernardo2001discrete,joshi1997local,vereschagin1996asymptotics} determined special solutions that are asymptotically pole free in sectors of the complex plane, and used transseries to study matrix models associated with discrete Painlev\'{e} equations \cite{aniceto2012resurgence}. 

We will explore the asymptotic behaviour of solutions to discrete Painlev\'{e} I in an important continuum limit which connects dPI with the continuous first Painlev\'{e} equation \cite{cresswell1999discrete}, in order to explain how the integrability of the system is linked to Stokes' phenomenon in this system. Our analysis will show that the solution for this special case does not contain Stokes switching, as the Stokes multipliers are equal to zero. As we shall see, this means that the solution to the corresponding infinite-order differential equation does not contain any moveable singularities, which is a marker of integrability in continuous problems.

Finally, we note that other exponential asymptotic methods, such as Borel transform analysis, have been used to study exponentially small terms in discrete equations, such as the hyperasymptotic analysis of second-order linear difference equations in \cite{daalhuis2004inverse}. It is likely that such methods could also be applied to the discrete problems considered in this study.

\subsection{Paper Outline}

In Section \ref{s:background}, we introduce the key equations that will be studied in this paper, and discuss their relevant properties. We first introduce the continuous Painlev\'{e} I equation, and the two infinite-order differential equations corresponding to the finite-difference discretisation and the first discrete Painlev\'{e} equation. We then introduce two families of differential equations, obtained by truncating the infinite-order differential equations at finite order $K$. 

In {Section \ref{s:FDK}}, we calculate the Stokes switching in members of the differential equation family obtained from the finite-difference discretisation truncated after $K$ terms. In Section \ref{s:FDinf}, we calculate the Stokes switching in the infinite-order differential equation, and compare this with the results from the preceding section to determine how the behaviour seen in the discrete Stokes' phenomenon arises as $K \to \infty$. 

In Section \ref{s:integrable}, we perform a similar analysis to Sections \ref{s:FDK} and \ref{s:FDinf} for the integrable discrete Painlev\'{e} I equation. We will see that, although the finite-order differential equations demonstrate Stokes' phenomenon, it disappears in the analysis of the discrete equation, ensuring that the only moveable singularities in the solution are poles. 

In Section \ref{s:GenDelta}, we consider a generalized problem that includes both the finite-difference and integrable discrete equations, and show that there are several other related systems that do not possess any exponentially small contributions to the asymptotic solution. We summarise the results and outline the key conclusions in Section \ref{s:conclusions}. A transseries analysis of the solutions is included in Appendix \ref{s:AppTrans} and \ref{s:AppTransGen}, which numerical descriptions of moveable singularities to the non-integrable equations in the complex plane.

\section{Background}\label{s:background}

\subsection{Painlev\'{e} I}

The first Painlev\'{e} {equation is} given by
\begin{equation}\label{e:P1}
\diff{^2u}{z^2} = 6u^2 + z, 
\end{equation}
where we consider the domain $z \in \mathbb{C}$. This equation possesses the Painlev\'{e} property; all singular points that depend on the initial data, known as moveable singularities, are poles \cite{flaschka1991integrability,ramani1989painleve}. This property is commonly used as a marker of integrability in ordinary differential equations. 

The first Painlev\'{e} equation \eqref{e:P1} has special solutions, known as tritronqu\'{e}e solutions, in which the poles are restricted to sectors of the complex plane. These solutions divide the plane up into five sectors with angle $2\pi/5$, and the {tritronqu\'{e}e} solutions contain poles in only one of the five sectors, identical up to rotation. Tritronqu\'{e} solutions have been studied extensively in numerous articles, for example \cite{joshi2001boutroux,novokshenov2010poles,bertola2013universality,deano2023riemann}.

\begin{figure}
\centering
\includegraphics{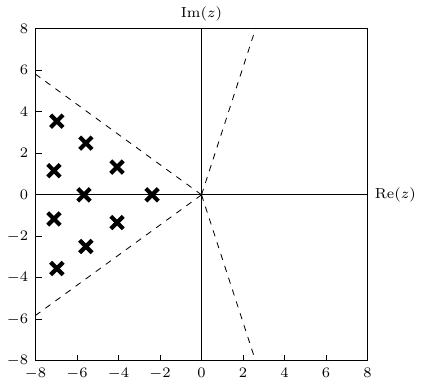}
\caption{Pole locations for a tritronqu\'{e}e solution of the Painlev\'{e} I equation \eqref{e:P1}. Poles are indicated by crosses, and are only found in the sector $\mathrm{arg}(z) \in (4\pi/5,6\pi/5)$. The five sectors of the solution are separated by dashed lines. Pole locations for the other {tritronqu\'{e}e} solutions may be found by rotating the pole locations {by a} multiple of $2\pi/5$.}\label{f:ttq_poles}
\end{figure}

If we require that the solution is algebraic in the limit that $z \to \infty$ on the real axis, then the asymptotic behaviour on this axis is given by
\begin{equation}
u = \i\sqrt{\frac{z}{6}} + \mathcal{O}\left(\frac{1}{z^{1/4}}\right), \qquad z \rightarrow \infty.
\end{equation}
This series describes the asymptotic behaviour of the tritronqu\'{e}e solution in the limit that $|z| \to \infty$ in the region $\mathrm{arg}(z) \in (-4\pi/5,4\pi/5)$. These sectors do not contain poles \cite{costin2014proof}. All of the poles in the solution are instead found within the sector $\mathrm{arg}(z) \in (4\pi/5, 6\pi/5)$, which contains the negative real axis. The first few pole locations are shown in Figure \ref{f:ttq_poles} for a typical tritronqu\'{e} solution, and approximate numerical values for these poles are given in Table \ref{t:novokshenov} \cite{novokshenov2010poles}. This special solution corresponds to the boundary conditions $u(0) \approx -0.187554308341189$ and $u'(0) \approx 0.304905560260202$ \cite{joshi2001boutroux}.

A local expansion of the solution near the poles gives the asymptotic behaviour near $z = z_p$ as
\begin{equation}\label{e:P1poles}
u \sim \frac{1}{(z-z_p)^2}, \qquad z \rightarrow z_p.
\end{equation}
For later algebraic convenience, we apply the transformation $y = -2u$. This gives the differential equation \eqref{e:P1m}. We select the solution which has asymptotic behaviour
	\begin{align}\label{e:P1m_a1}
y &\sim \i\sqrt{\frac{2z}{3}} \quad \mathrm{as} \quad z \rightarrow \infty. \\
y &\sim -\frac{2}{(z-z_p)^2}\quad \mathrm{as} \quad z \rightarrow z_p,\label{e:P1m_a2}
\end{align}
where $z_p$ are the same pole locations used to describe the describe the asymptotic behaviour in \eqref{e:P1poles}.

\begin{table}
\begin{center}
\begin{tabular}{ |c c| } 
 \hline
Pole Index & Location \\ 
 \hline
 $z_0$ &  $-2.384168769564550$  \\ 
$z_{1,2}$ &  $-4.071055433084488 \pm 1.3355514706611683\i$  \\ 
 $z_3$ &  $-5.664449429020133$  \\ 
$z_{4,5}$ &  $-5.573539567139687 \pm 2.489136512145025\i$  \\ 
$z_{6,7}$ &  $-6.971620941710831 \pm 3.548265719706162\i$  \\ 
$z_{8,9}$ &  $-7.110821873372145 \pm 1.162265719858028\i$  \\ 
 \hline
\end{tabular}
\end{center}
\caption{The nearest poles to the origin in the tritronqu\'{e}e solution to Painlev\'{e} I \cite{novokshenov2010poles}, shown in Figure \ref{f:ttq_poles}.}
\label{t:novokshenov}
\end{table}

\subsection{Finite-Difference Discretisation}\label{s:FD}

We can apply a straightforward finite-difference discretisation to \eqref{e:P1m}. Define the finite-difference step to be $\delta z = \epsilon$. Using a second-order central difference stencil to approximate the second derivative  in \eqref{e:P1m} gives
\begin{equation}\label{e:FD}
\frac{y(z - \epsilon) - 2 y(z) + y(z+\epsilon)}{\epsilon^2} + 3 y(z)^2 = -2z.
\end{equation}
Formally expanding this expression around $\epsilon = 0$ gives
\begin{equation}\label{e:FDfam}
2\sum_{m=1}^{\infty}\frac{\epsilon^{2m-2}}{(2m)!} \diff{^{2m}y(z)}{z^{2m}} + 3y(z)^2 = -2z.
\end{equation}
{This provides a continuous interpolation $y(z)$ between the discrete solution points. This interpolation is not unique; for example, it is always possible to add multiples of $\sin(2\pi z/\eps)$ to $y(z)$ to produce a new function that takes the same value at each finite difference step $z = z_0 + n \eps$ for $n \in \mathbb{Z}$. These terms oscillate rapidly in the limit that $\eps \to 0$, meaning that an interpolation containing these terms is likely to be highly singular in the asymptotic limit. The interpolation generated by \eqref{e:FD} does not contain such highly singular oscillations.}

Motivated by the form of the {tritronqu\'{e}e} solution, we would like to pose the condition that on the positive real axis, $y(z) = \mathcal{O}(\sqrt{z})$ as $z \to \infty$. In fact, this condition does not uniquely define the solution, as there is an exponentially-small term with an arbitrary constant present in the solution on the positive real axis. Instead, we prescribe the condition that $y(z) = \mathcal{O}(\sqrt{z})$ as $|z| \to \infty$ for $\mathrm{Re}(z) > 0$. This forces the arbitrary constant to be zero in a region containing the positive real axis and therefore does uniquely specify the solution to the differential equation.

We define the asymptotic power series
\begin{equation}\label{e:dPIseries}
y(z) \sim \sum_{j=0}^{\infty} \epsilon^{2j}y_j(z)\qquad \mathrm{as} \qquad \epsilon \to 0.
\end{equation}
The far-field condition gives $y_0 = \mathcal{O}(\sqrt{z})$ as $|z| \to \infty$ in the right half-plane. The leading-order behaviour $y_0$ satisfies the differential equation \eqref{e:P1m}.  We select $y_0$ to be the {tritronqu\'{e}e} solution, as the asymptotic behaviour as $|z| \to \infty$ is given by \eqref{e:P1m_a1}, which is consistent with the far-field condition for $y_0$. This solution for $y_0$ contains double poles at the locations in Table \ref{t:novokshenov}, the asymptotic behaviour near these poles being given by \eqref{e:P1m_a2}.

\subsection{Discrete Painlev\'{e} I}\label{s:dP1}

An important discrete equation is the discrete first Painlev\'{e} equation, or discrete Painlev\'{e} I, {given by}
\begin{equation}\label{e:dP1}
w_{n+1} + w_n + w_{n-1} = \frac{c_1 + c_2 n}{w_n} + c_3,
\end{equation}
where $c_j$ are constants for $j = 1, 2, 3$. We consider a distinguished continuum limit {from \cite{cresswell1999discrete}}, whereby
\begin{equation}\label{e:cont}
z = \epsilon n ,\qquad w_n = 1 + \epsilon^2 y(z),
\end{equation}
with
\begin{equation}
c_1 = -3 + r_1 \epsilon^2, \qquad c_2 = r_2 \epsilon^5, \qquad c_3 = 6 - r_1 \epsilon^2 - r_3 \epsilon^4,
\end{equation}
where $r_j$ are constants for $j = 1,2,3$.  This gives 
\begin{align}\nonumber
(1 + \epsilon^2 y(z))(3 + \epsilon^2 y(z + \epsilon) & + \epsilon^2 y(z) + \epsilon^2 y(z-\epsilon))  \\ & 
\label{e:dP1c}
= 3 + r_2 \epsilon^4 z + 6 \epsilon^2 y(z) - r_1 \epsilon^4 y(z) - r_3 \epsilon^4 (1 + y(z)).
\end{align}
Formally expanding about $\epsilon=0$ gives 
\begin{align}\nonumber
2 \epsilon^2 \sum_{m=1}^{\infty} \frac{\epsilon^{2m}}{(2m)!} \diff{^{2m}y(z)}{z^{2m}} + 3 \epsilon^4 y(z)^2 +  2 \epsilon^4 y(z) & \sum_{m=1}^{\infty} \frac{\epsilon^{2m}}{(2m)!} \diff{^{2m}y(z)}{z^{2m}}  \\ & \label{e:dP1t}
= r_2 \epsilon^4 z - r_1 \epsilon^4 y(z) - r_3 \epsilon^4(1+ \epsilon^2 y(z)). 
\end{align}
As in the finite-difference case, we prescribe the condition that $y(z) = \mathcal{O}(\sqrt{z})$ as $|z| \to \infty$ for $\mathrm{Re}(z) > 0$. We define the asymptotic power series \eqref{e:dPIseries}, and we substitute this into \eqref{e:dP1t}. The leading-order behaviour in the limit that $\epsilon \to 0$ is given by balancing at $\mathcal{O}(\epsilon^4)$. This behaviour is governed by
\begin{equation}
\diff{^2 y_0}{z^2} + 3 y_0(z)^2 = r_2 z - r_1 y(z) - r_3.
\end{equation}
This equation can always be transformed into the Painlev\'{e} I equation \eqref{e:P1} using a suitable change of variables \cite{cresswell1999discrete}. We simplify the process by selecting $r_2 = -2$ and $r_1 = r_3 = 0$, giving
\begin{equation}\label{e:scaleddP1}
\diff{^2 y_0}{z^2} + 3 y_0(z)^2 = -2 z,
\end{equation}
so that $y_0$ satisfies \eqref{e:P1m}. The far-field boundary condition implies that $y_0(z) = \mathcal{O}(\sqrt{z})$ as $|z| \to \infty$ in the right half-plane. This condition permits us to again choose $y_0$ to be the {tritronqu\'{e}e} solution, which has the asymptotic behaviour in \eqref{e:P1m_a1} as $z\to \infty$. We again note that $y_0$ contains double poles at the locations in Table \ref{t:novokshenov} and the local asymptotic behaviour near the poles is given by \eqref{e:P1m_a2}.

\subsection{Ordinary Differential Equation Families}

The two difference equations described in Section \ref{s:FD} and \ref{s:dP1} both tend to the scaled continuous Painlev\'{e} I equation \eqref{e:P1m} in an appropriate continuum limit. The difference equation \eqref{e:FD} is not integrable, while \eqref{e:dP1} is. In both cases the continuum limit is singular, as can be seen from the infinite-order differential equations \eqref{e:FDfam} and \eqref{e:dP1t}. Because these are singularly perturbed limits, we expect Stokes' phenomenon to appear in the asymptotic behaviour of the solutions. We aim to compare the behaviour of Stokes' phemenon in the {integrable} \eqref{e:dP1t} and non-integrable \eqref{e:FDfam} discrete equations.

In order to understand more clearly how Stokes' phenomenon in the discrete equations \eqref{e:FD} and \eqref{e:dP1t} arises in the limit that $\epsilon \to 0$, we consider two families of ordinary differential equations indexed by $K \in \mathbb{Z}$ with $K \geq 1$. These families are obtained by truncating the series terms in \eqref{e:FD} and \eqref{e:dP1t} so that the highest power of $\epsilon$ present in the truncated version of \eqref{e:dP1t} is $\epsilon^{2K+2}$. 

After some simplification, the first family is given by
\begin{align}\label{e:familyFD}
2  \sum_{m=1}^{K} \frac{\epsilon^{2m-2}}{(2m)!} \diff{^{2m}y(z)}{z^{2m}} + 3 y(z)^2  = -2 z.
\end{align}
The second family is given by
\begin{align}\label{e:family}
2  \sum_{m=1}^{K} \frac{\epsilon^{2m-2}}{(2m)!} \diff{^{2m}y(z)}{z^{2m}} + 3 y(z)^2 +  2 y(z)  \sum_{m=1}^{K-1} \frac{\epsilon^{2m}}{(2m)!} \diff{^{2m}y(z)}{z^{2m}}  = -2 z.
\end{align}
We see that the second equation family contains an extra term that causes the equation in the limit that $K \to \infty$ to be integrable. In both cases, balancing at leading order in the limit $\epsilon \to 0$ shows that the leading-order behaviour is governed by \eqref{e:P1m}. As $K \to \infty$, we obtain infinite-order differential equations corresponding to the finite-difference Painlev\'{e} I equation \eqref{e:FDfam} and the discrete Painlev\'{e} I equation \eqref{e:dP1t}.

Both \eqref{e:familyFD} and \eqref{e:family} can be written in the more general form
\begin{align}\label{e:familygen}
2  \sum_{m=1}^{K} \frac{\epsilon^{2m-2}}{(2m)!} \diff{^{2m}y(z)}{z^{2m}} + 3 y(z)^2 +  (2 + \Delta) y(z)  \sum_{m=1}^{K-1} \frac{\epsilon^{2m}}{(2m)!} \diff{^{2m}y(z)}{z^{2m}}  = -2 z,
\end{align}
where $\Delta = 0$ gives the equation family \eqref{e:family} that tends to discrete Painlev\'{e} I as $K \to \infty$, and the family \eqref{e:familyFD} that gives the finite-difference discretisation as $K \to \infty$ corresponds to $\Delta = -2$. 

{In Section \ref{s:GenDelta}, we will allow $\Delta$ to vary in order to show that the Stokes constants in the problem change continuously with $\Delta$. 
\begin{align}\label{e:Delta1}
\left(\tfrac{1}{\eps^2} + \left(1 + \tfrac{\Delta}{2}\right)w_n\right)(w_{n+1}-2w_n + w_{n-1}) + 3 w_n^2  = -2 \epsilon n.
\end{align}}
We will also find in Section \ref{s:GenDelta} that the $\Delta = 1$ case is of interest, as the discrete equation may be evaluated exactly.

\section{Finite-Difference Family: Finite \texorpdfstring{$K$}{K}}\label{s:FDK}

The first few equations in the finite-difference family \eqref{e:familyFD} are given by
\begin{align}
\label{e:1}
K &= 1: \qquad \diff{^2y}{z^2} + 3y^2  = -2z,\\
\label{e:2}
K &= 2: \qquad \frac{\epsilon^2}{12}\diff{^4y}{z^4} + \diff{^2y}{z^2} + 3y^2  = -2z,\\
\label{e:3}
K&=3: \qquad  \frac{\epsilon^4}{360}\diff{^6y}{z^6}+ \frac{\epsilon^2}{12}\diff{^4y}{z^4} +\diff{^2y}{z^2} + 3y^2  = -2z.
\end{align}
The $K=1$ case is not singularly perturbed, but is simply the continuous Painlev\'{e} I equation \eqref{e:P1m}. 

We will present the Stokes structure of the $K=2$ and $K=3$ cases to demonstrate the two classes of behaviour, $K$ even and $K$ odd, that can be seen in this family. We will then explain the Stokes structure that appears in solutions to the entire family of differential equations generated by \eqref{e:familyFD}.

\subsection{Series Terms}																															

By substituting the power series expansion from \eqref{e:dPIseries} into {\eqref{e:familyFD}}, we obtain
\begin{align}
\sum_{m=1}^K \sum_{j=0}^{\infty} \frac{2\epsilon^{2m+2j-2}}{(2m)!}\diff{^{2m} y_j}{z^{2m}} + 3 \sum_{j=0}^{\infty}\sum_{l=0}^{\infty} \epsilon^{2j+2l}y_j y_l     = -2 z,
\end{align}
where $y_j$ is a function of $z$. 

Balancing at $\mathcal{O}(1)$ as $\epsilon \to 0$, we obtain the leading order equation \eqref{e:P1m}, and choose the tritronqu\'{e}e solution with asymptotic behaviour described in \eqref{e:P1m_a1}--\eqref{e:P1m_a2} as the leading-order solution $y_0$. This solution contains poles in the region $\mathrm{arg}(z) \in (4\pi/5,6\pi/5)$; if the leading-order solution of a singularly-perturbed equation contains singular points, these points generally produce Stokes lines and anti-Stokes lines in the full asymptotic solution {\cite{Chapman, Dingle}}. Balancing terms at $\mathcal{O}(\epsilon^{2k})$ with $k > K$ in the limit that $\epsilon \to 0$ gives for $k \geq 1$ the recurrence relation
\begin{align}\label{e:recur}
\sum_{m=1}^K\frac{2}{(2m)!}\diff{^{2m-2} y_{k-m+1}}{z^{2m-2}} + 3\sum_{l=0}^{k} y_l y_{k-l} = 0, 
\end{align}
which can be used to obtain $y_{k}$ in terms of earlier series coefficients. This is a linear ordinary differential equation in terms of $y_{k}$, an therefore does not introduce new singular behaviour. Hence, $y_k$ is singular at pole locations of the {tritronqu\'{e}e} solution to Painlev\'{e} I \eqref{e:dP1}. 

For $K=2$ and $K=3$, \eqref{e:recur} gives
\begin{align}\label{e:recur2}
K = 2 :&\qquad \frac{1}{12}\diff{^4 y_{k-1}}{z^4} + \diff{^2 y_{k}}{z^2}  + 3\sum_{l=0}^{k} y_l y_{k-l} = 0.\\
\label{e:recur3}
K = 3 :&\qquad \frac{1}{360}\diff{^6 y_{k-2}}{z^6} + \frac{1}{12}\diff{^4 y_{k-1}}{z^4} + \diff{^2 y_{k}}{z^2}  + 3\sum_{l=0}^{k} y_l y_{k-l} = 0.
\end{align}

\subsection{Late-Order Terms}
We assume the late-order terms satisfy a factorial-over-power ansatz of the form
\begin{equation}\label{e:ansatz}
y_k(z) \sim \frac{Y(z)\Gamma(2k + \gamma)}{\chi(z)^{2k+\gamma}} \quad \mathrm{as} \quad k \to \infty
\end{equation}
{for some constant parameter $\gamma$ to be determined.} To find the late-order terms for \eqref{e:2}, corresponding to $K=2$, we substitute the ansatz \eqref{e:ansatz} into \eqref{e:recur}. In the limit that $k \to \infty$, using the fact that $k+1 > K$ for any fixed $K$ in the limit, we find that
\begin{align}\nonumber
&\left[\sum_{m=1}^K \frac{2\chi'(z)^{2m}}{(2m)!}\right]\frac{Y(z)\Gamma(2k+\gamma+2)}{\chi(z)^{2k+\gamma+2}} \\& - \left[\sum_{m=1}^K \frac{2\chi'(z)^{2m-1} Y'(z)}{Y(z)(2m-1)!} + \frac{\chi'(z)^{2m-2} \chi''(z)Y(z)}{(2m-2)!} \right]\frac{Y(z)\Gamma(2k+\gamma+1)}{\chi(z)^{2k+\gamma+1}}  + \ldots = 0,\label{e:laterecur}
\end{align}
where the omitted terms are {$\mathcal{O}(y_k)$} in the limit that $k \to \infty$. By balancing at leading order in the limit that $k \to \infty$, we find that the equation for the singulant associated with the pole at $z = z_p$ is therefore given, {after division by $\chi'(z)^2$}, by
\begin{equation}\label{e:singulantK}
\sum_{m=1}^K \frac{2\chi'(z)^{2m-2}}{(2m)!}=0, \qquad \chi(z_p) = 0.
\end{equation}
This has $2K-2$ nonzero constant solutions for $\chi'$. We denote these solutions by $\chi_1$, $\chi_2$, \ldots, $\chi_{2K-2}$. Note that we will later perform an analysis corresponding to $K = \infty$ in \eqref{e:singinfext} as part of the analysis of the discrete finite-difference problem.

Balancing \eqref{e:laterecur} to the next order as $k \to \infty$ gives $Y'(z) = 0$. We denote this as $Y(z) = \Lambda$, where $\Lambda$ is a constant that will later be determined by balancing the late-order terms \eqref{e:ansatz} with an inner expansion of $y(z)$ in the neighbourhood of $z = z_p$. 

In order for the late-order terms to be consistent with the asymptotic behaviour of the leading-order solution near the poles from \eqref{e:P1m_a2} at $z = z_p$, we require that the late-order expression be singular with singularity strength 2 when $k$ is set to zero, giving $\gamma = 2$. 

To determine $\Lambda$, we must balance the late-order ansatz \eqref{e:ansatz} with a local expansion of $y(z)$ near $z = z_p$. The series expansion \eqref{e:dPIseries} is not well-ordered in the asymptotic limit if $z - z_p = \mathcal{O}(\epsilon)$. {It therefore ceases to be valid in the regime where $\epsilon(z-z_p)^{-1}$ is not small in the limit $\epsilon \to 0$, i.e. when $z - z_p = \mathcal{O}(\epsilon)$. This motivates the rescaling of} \eqref{e:familyFD} in terms of the rescaled variables 
\begin{equation}\label{e:innervar}
z - z_p = \epsilon \eta, \qquad y(z) = \frac{v(\eta)}{\epsilon^2}.
\end{equation}
This change of variables gives
\begin{equation}\label{e:innerK}
{2}\sum_{m=1}^{K} \frac{1}{(2m)!}\diff{^{2m}v(\eta)}{\eta^{2m}} + {3 v(\eta)^2}= -2{\epsilon^4} (z_p + \epsilon \eta).
\end{equation}
To describe the behaviour of the solution near the singularity $z_p$, we require only the leading-order terms as $\epsilon \to 0$, and therefore neglect the right-hand side. We expand the leading-order inner solution as
\begin{equation}\label{e:innerserK}
v(\eta) \sim \sum_{j = 0}^{\infty}\frac{v_j}{\eta^{2j+2}}\quad \mathrm{as} \quad |\eta|\to \infty,
\end{equation}
where $v_0 = -2$ is required to be consistent with the asymptotic behaviour \eqref{e:P1m_a2}. Substituting this expression into \eqref{e:innerK} and neglecting the subdominant terms in the limit that $\eps \to 0$ gives 
\begin{equation}\label{e:innerFD}
2\sum_{m=1}^{K}\frac{1}{(2m)!}\sum_{j=0}^{\infty}\frac{(2j+2m+1)!}{(2j+1)!}\frac{v_j}{\eta^{2j+2m+2}} + 3\sum_{\ell = 0}^{\infty}\sum_{j=0}^{\infty}\frac{v_{\ell}v_j}{\eta^{2j + 2\ell + 4}} = 0.
\end{equation}
Balancing this expression at {$\mathcal{O}(\eta^{-2k-4})$}, we obtain
\begin{equation}\label{e:finiteKrecur}
2\sum_{m=1}^{\mathrm{min}(K,k+1)}\binom{2k+3}{2m}v_{k-m+1} + 3 \sum_{\ell = 0}^{k}v_{\ell}v_{k-\ell} = 0.
\end{equation}
This gives the recurrence relation for $v_k$
\begin{equation}
2(k+3)(2k-1)\,v_k = 2\sum_{m=2}^{\mathrm{min}(K,k+1)}\binom{2k+3}{2m}v_{k-m+1} + 3\sum_{\ell=1}^{j-1}v_{\ell}v_{k-\ell}.
\end{equation}
Using this recurrence relation, values of $v_j$ may be calculated to arbitrarily large order. Note that the number of terms in the first sum in the recurrence relation increases until $j > K-1$, beyond which the recurrence relation does not change. This will not be true for the discrete equation (i.e. $K \to \infty$), in which the summation term grows indefinitely.

The final step is to match the inner series \eqref{e:innerserK} with the late-order ansatz \eqref{e:ansatz}. For simplicity, we will demonstrate the matching explicitly for $K=2$ in Section \ref{S:singK2}, and will later contrast the process for obtaining this result with the equivalent analysis for the discrete problem (i.e. $K \to \infty$) in both the finite-difference and integrable cases.

The contribution to $y_k$ as $k \to \infty$ from the pole at $z=z_p$ is given by a sum of terms with the form
\begin{equation}\label{e:LOT3}
    y_k \sim \sum_{\ell = 1}^{2K-2}\frac{\Lambda_{\ell} \Gamma(2k+2)}{\chi_{\ell}^{2k+2}}.
\end{equation}
From this, we are able to determine the Stokes structure of the solution caused by the pole at $z = z_p$.

\subsubsection{\texorpdfstring{$K=2$}{K=2}}\label{S:singK2}

\begin{figure}[tbp]
\centering
\includegraphics{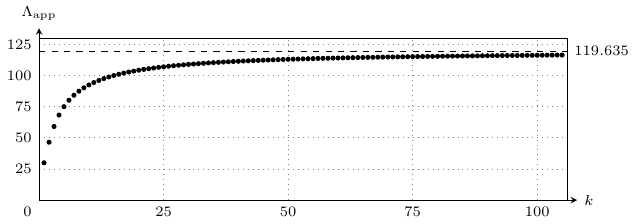}
\caption{Approximation of $\Lambda$ obtained by evaluating \eqref{e:K2match} at different values of $k$, denoted $\Lambda_{\mathrm{app}}$. The value of $\Lambda_{\mathrm{app}}$ obtained as $k$ becomes large is shown as a dashed line.}\label{fig:K}
\end{figure}

For the case $K=2$, the singulant equation becomes
\begin{equation}\label{e:singulant2}
\frac{\chi'(z)^2}{12} + 1 = 0,\qquad \chi(z_p) = 0.
\end{equation}
Each pole produces two singulants,
\begin{equation}\label{e:singulant2exp}
\chi_1 = 2\i\sqrt{3}(z-z_p), \qquad \chi_2 = -2\i\sqrt{3}(z-z_p).
\end{equation}
Note that $\chi'$ is imaginary here. In any case where $K$ is even, we will obtain at least one pair of imaginary values for $\chi'$, due to the symmetry of solutions to \eqref{e:singulantK}. Each of the two singulants is associated with a corresponding value of $\Lambda_{\ell}$ for $\ell = 1,2$. By expressing the terms of the inner series in terms of the outer variable $z$, we find that setting $j = 2k$ gives the matching condition
\begin{equation}\label{e:prefactormatchK2}
\frac{v_{k}}{(z - z_p)^{{2k}+2}} \sim  \frac{(\Lambda_{1}+\Lambda_2)\Gamma(2k+2)}{[2\i\sqrt{3}(z - z_p)]^{2k+2}} \quad \mathrm{as} \quad k \to \infty.
\end{equation}
The ansatz \eqref{e:ansatz} is chosen under the assumption that there are no terms in the expansion with odd powers of $\epsilon$. This condition requires that $\Lambda_1$ and $\Lambda_2$ be identical; if this were not the case, we would have to include terms with odd powers of $\epsilon$ in \eqref{e:innerserK}. For simplicity, we subsequently denote both $\Lambda_1$ and $\Lambda_2$ as $\Lambda$. The matching condition \eqref{e:prefactormatchK2} therefore gives
\begin{equation}\label{e:K2match}
2\Lambda= \lim_{k \to \infty} \frac{v_{k}(2\i\sqrt{3})^{2k+2}}{\Gamma(2k+2)}.
\end{equation}
The value of the prefactor constants can be approximated by calculating $v_{2k}$ up to some large value of $k$, and evaluating the right-hand side of this expression. In Figure \ref{fig:K}, we illustrate the value of $\Lambda$ obtained by evaluating this expression for various $k$, denoted $\Lambda_{\mathrm{app}}$, illustrating the convergence of the approximation to a constant for larger values of $k$. Evaluating this expression at $k = 2000$ gives $\Lambda \approx 119.635$. Hence, 
\begin{equation}\label{e:LOT3n}
    y_k \sim \frac{\Lambda \Gamma(2k+2)}{[2\i\sqrt{3}(z - z_p)]^{2k+2}} + \frac{\Lambda \Gamma(2k+2)}{[-2\i\sqrt{3}(z - z_p)]^{2k+2}}, \qquad \Lambda \approx 119.635.
\end{equation}

\subsubsection{\texorpdfstring{$K=3$}{K=3}}\label{S:singK3}

For the case $K=3$, the singulant equation becomes
\begin{equation}\label{e:singulant3}
\frac{\chi'(z)^4}{360} + \frac{\chi'(z)^2}{12} + 1 = 0,\qquad \chi(z_p) = 0.
\end{equation}
In this case, each pole produces four singulants, which we will subsequently denote by
\begin{align}\chi_{1} = \sqrt{-15-3\i\sqrt{15}}(z - z_p), \qquad \chi_{2} = -\sqrt{-15-3\i\sqrt{15}}(z-z_p),\\
\chi_{3} = \sqrt{-15+3\i\sqrt{15}}(z - z_p), \qquad \chi_{4} = -\sqrt{-15+3\i\sqrt{15}}(z-z_p).\label{e:allsingulants}
\end{align}
In this case there are no imaginary values of $\chi'(z)$. This is not necessarily true for all odd values of $K$, and we will later show that as $K$ grows, we find solutions to the singulant equation for odd $K$ that are arbitrarily close to the imaginary axis. The constants $\Lambda_{\ell}$ for $\ell = 1,\ldots,4$ can be calculated in similar fashion to Section \ref{S:singK2}, but we do not require that result for the present analysis.

\subsection{Stokes' Phenomenon}

Once we have the form of the late-order terms \eqref{e:LOT3}, we may determine the Stokes structure of the solutions. Stokes lines are curves across which exponentially small terms experience rapid jumps, often switching on or off entirely. From \cite{Dingle}, we know that Stokes lines corresponding to exponentially small terms that are switched by an algebraic power series satisfy the condition
\begin{equation}
\mathrm{Im}(\chi) = 0, \qquad \mathrm{Re}(\chi) > 0.\label{e:SLcond}
\end{equation}
Anti-Stokes lines are curves across which exponential terms change between being exponentially small to exponentially large in the asymptotic limit. These curves satisfy the condition
\begin{equation}
\mathrm{Re}(\chi) = 0.\label{e:ASLcond}
\end{equation}
It is important to note that, given we are dealing with nonlinear problems, in the presence of exponentially large terms we can no longer describe the solution behaviour using the power series \eqref{e:dPIseries} beyond these curves. The exponential terms in the solution will instead grow, and the dominant asymptotic balance in the equation will change.

\subsubsection{\texorpdfstring{$K=2$}{K=2}}

If $K=2$ we find that there are two exponential contributions associated with each leading-order pole, given by $\chi_1$ and $\chi_2$ in \eqref{e:singulant2exp}. We will refer to the corresponding exponential contributions as $u_{\mathrm{exp},1}$ and $u_{\mathrm{exp},2}$ respectively. The algebraic form of these contributions is calculated in Section \ref{s:FDstokes}.

Stokes lines originate at leading-order singularities. We consider the Stokes lines originating at $z = z_p$ due to the leading-order pole. The condition \eqref{e:SLcond} shows that the Stokes line associated with $\chi_1$  extends vertically upwards from $z = z_p$, while the Stokes line associated with $\chi_2$ extends vertically downwards. From the condition \eqref{e:ASLcond}, we conclude that anti-Stokes lines for both contributions extend horizontally from $z = z_p$.

\begin{figure}
\centering
\subfloat[Horizontal cut]{
\includegraphics{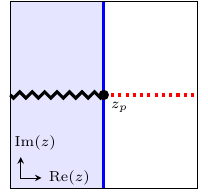}
}
\subfloat[Upwards cut]{
\includegraphics{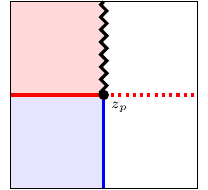}
}
\subfloat[Downwards cut]{
\includegraphics{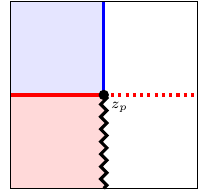}
}

\subfloat[Stokes structure of $y(z)$ with $K=2$]{
\includegraphics{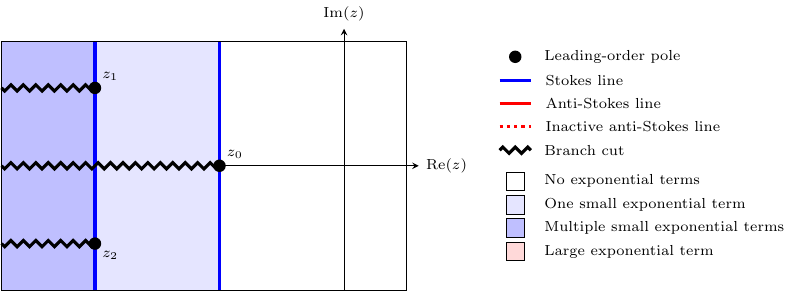}
}

\caption{The diagrams in (a)--(c) show local Stokes' phenomenon in $y(z)$ for $K=2$ generated by a singularity at $z = z_p$, with the branch cut set (a) horizontally, (b) vertically upwards, and (c) vertically downwards. The legend in each case is shown in (d). Figure (d) shows the combined Stokes structure due to multiple poles. Each pole generates vertical Stokes lines which switch on exponentially small contributions to the left of the Stokes line. There are no active anti-Stokes lines, so the series expansion \eqref{e:dPIseries} is valid everywhere except on the branch cuts. {There are no exponential contributions in the right-half plane (and hence the rightward extending anti-Stokes lines are inactive), as this Stokes structure corresponds to the unique asymptotic solution which is $\mathcal{O}(\sqrt{z})$ as $|z| \to \infty$ for $\mathrm{Re}(z) > 0$.}}\label{fig:SL2}
\end{figure}

Recall that the boundary condition on the differential equation requires that the solution be $\mathcal{O}(\sqrt{z})$ as $|z| \to \infty$ for $\mathrm{Re}(z) > 0$. However, both $u_{\mathrm{exp},1}$ and $u_{\mathrm{exp},2}$ are exponentially large in some region to the right of $z=z_p$, due to the anti-Stokes line extending horizontally from the pole: the contribution $u_{\mathrm{exp},1}$ is exponentially large below the anti-Stokes line, while the contribution $u_{\mathrm{exp},2}$ is exponentially large above it. The only way this can be prevented is if both contributions are inactive on the right-hand side of the pole, or $\mathrm{Re}(z) > \mathrm{Re}(z_p)$. This configuration is illustrated in Figure \ref{fig:SL2}(a). Note that this Stokes structure requires that the solution contain a branch cut, which we will set to extend horizontally leftwards from the pole. 

Note that we could set the branch cuts to extend vertically, as in Figure \ref{fig:SL2}(b) and (c). In this case, the exponentially small contribution would reach the anti-Stokes line following $\mathrm{Im}(z) = \mathrm{Im}(z_p)$, and there would be a region in which the exponential terms become large and change the dominant balance of the equation, invalidating the original expression. These regions are not present on the principal Riemann sheet if we set the branch cuts to be horizontal\footnote{Such comments already highlight the exceptional status of the integrable case, in which branch points are absent.}.

The Stokes structure depicted in Figure \ref{fig:SL2}(a) will be repeated for each pole, introducing new exponential contributions and branch cuts originating at each pole location. The combined exponential behaviour is shown in Figure \ref{fig:SL2}(d). Each pole generates two exponential contributions which appear across Stokes lines that extend vertically from the poles.

\subsubsection{\texorpdfstring{$K=3$}{K=3}}

If $K=3$ there are instead four exponential contributions associated with each leading-order pole, given by $\chi_1$ to $\chi_4$ in \eqref{e:allsingulants}. The corresponding Stokes and anti-Stokes lines originate at $z = z_p$. The conditions from \eqref{e:SLcond} and \eqref{e:ASLcond} now show that the Stokes and anti-Stokes lines satisfy
\begin{equation}\label{e:stokes3}
\frac{\mathrm{Re}(z-z_p)}{\mathrm{Im}(z-z_p)} = \frac{\pi}{2}\pm\tan\left(\frac{1}{2}\tan\left(\sqrt{\frac{3}{5}}\right)\right),
\end{equation}
and 
\begin{equation}\label{e:antistokes3}
\frac{\mathrm{Re}(z-z_p)}{\mathrm{Im}(z-z_p)} = \pm\tan\left(\frac{1}{2}\tan\left(\sqrt{\frac{3}{5}}\right)\right),
\end{equation}
respectively, where $\chi_1$ and $\chi_2$ select the upper sign and $\chi_3$ and $\chi_4$ select the lower sign. 

\begin{figure}
\centering
\subfloat[Contributions 1 and 2]{
\includegraphics{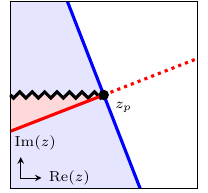}
}
\subfloat[Contributions 3 and 4]{
\includegraphics{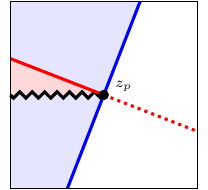}
}
\subfloat[All four contributions]{
\includegraphics{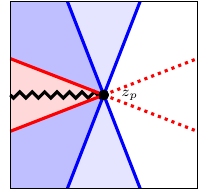}
}

\subfloat[Stokes structure of $y(z)$ with $K=3$]{
\includegraphics{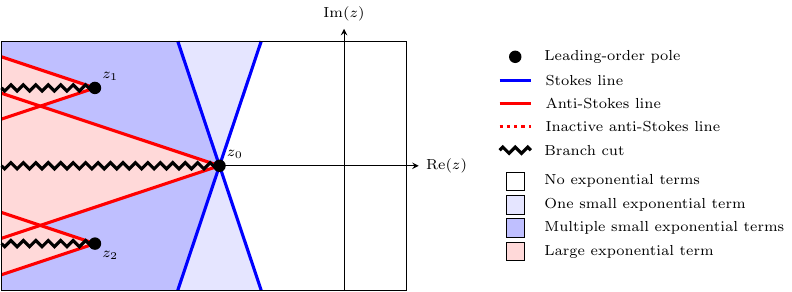}
}
\caption{The diagrams in (a)--(c) show local Stokes' phenomena in $y(z)$ for $K=3$ generated by a singularity at $z = z_p$ associated with (a) $\chi_1$ and $\chi_2$, (b) $\chi_3$ and $\chi_4$, and (c) the combined contribution of all four contributions.  The legend for each is shown in (d). Figure (d) shows the combined Stokes structure due to multiple poles. Each pole generates Stokes lines that switch on exponentially small contributions; these contributions become large when anti-Stokes lines are crossed. The series expansion \eqref{e:dPIseries} is no longer valid in the red-shaded regions.}\label{fig:SL3}
\end{figure}

The boundary condition requires that the solution not be exponentially large on the right-hand side of the poles, which implies that the exponential terms are absent on the right-hand side before appearing as the relevant Stokes lines are crossed from right to left. The Stokes structure associated with $u_{\mathrm{exp},1}$ and $u_{\mathrm{exp},2}$ is shown in Figure \ref{fig:SL3}(a), and the Stokes structure associated with $u_{\mathrm{exp},3}$ and $u_{\mathrm{exp},4}$ is shown in Figure \ref{fig:SL3}(b). In both cases, there is an anti-Stokes line on the left-hand side of the pole that causes an exponential term to become asymptotically large, and the expansion \eqref{e:dPIseries} then no longer describes the solution behaviour.

The combined effect of all four exponentials for a single pole is shown in Figure \ref{fig:SL3}(c). We see that there are exponentially small terms present in sectors above and below the pole, and a narrow sector to the left of the pole in which exponential terms become large, invalidating the original ansatz. The combined exponential behaviour is shown in Figure \ref{fig:SL2}(d).

\subsection{Stokes Switching}\label{s:FDstokes}

To determine the form of the exponential contribution that appears as Stokes lines are crossed, we apply exponential asymptotic techniques from \cite{Daalhuis,Chapman} for arbitrary $\chi$ and $K$. As this is a standard application of the method, we defer the details to Appendix \ref{s:SwitchFDK}. This analysis shows that the asymptotic quantity that is switched on across the Stokes line satisfying $\mathrm{Im}(\chi) = 0$ and $\mathrm{Re}(\chi) > 0$ is given by
\begin{equation}\label{e:RNapp}
   R_N \sim \frac{2\pi\i \Lambda}{\mu \epsilon^{2}} \e^{-\chi/\epsilon},\qquad \mathrm{where} \quad\mu = \sum_{m=1}^{K}\frac{2(2K)!}{(2m-1)!(\chi')^{2K-2m}}.
\end{equation}
Hence, each blue region in Figures \ref{fig:SL2}, \ref{fig:SL3}, and \ref{fig:stokesK} contains one or more exponentially small asymptotic contributions with this form. 

This description of the exponentially small behaviour is valid up to the anti-Stokes lines, across which the asymptotic series is no longer valid (depicted as red regions in Figure \ref{fig:SL3}). Calculating the behaviour of the exponential contributions beyond the anti-Stokes line is beyond the scope of the main text; however, a transseries analysis of the solution in regions where the exponential terms are large is included in Appendix \ref{s:AppTrans} for $K=3$ and Appendix \ref{s:AppTransGen} for some larger values of $K$. That analysis provides numerical evidence that the interaction between exponentials produces moveable branch cuts in the solution in these regions, which is expected behaviour in non-integrable  equations.

\subsection{Stokes structure for $K=4$, $K=5$}

In order to identify how the Stokes structure changes as $K$ increases, we examine solutions to the singulant equation \eqref{e:singulantK} for larger values of $K$, and the corresponding Stokes structure around leading-order poles at some $z = z_p$. Increasing the value of $K$ by one has the effect of increasing the number of exponential contributions by two. 

\begin{figure}
\begin{center}
\subfloat[$K=2$]{
\includegraphics{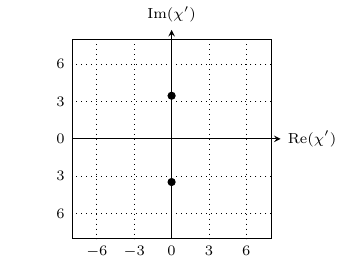}
}
\subfloat[$K=3$]{
\includegraphics{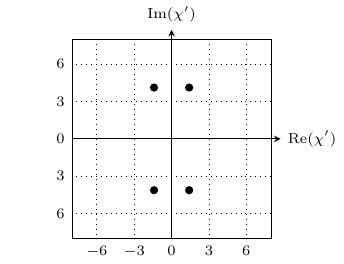}
}

\subfloat[$K=4$]{
\includegraphics{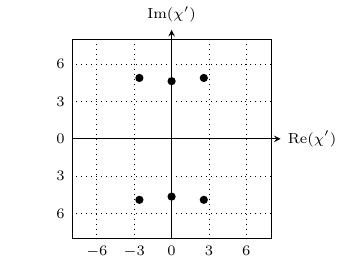}
}
\subfloat[$K=5$]{
\includegraphics{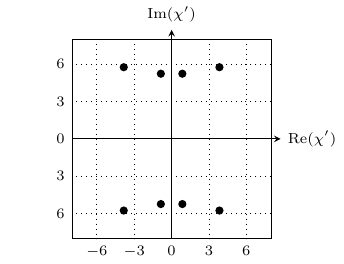}
}
\caption{Values of $\chi'$ for different choices of $K$, denoted by circles in the complex $z$-plane. Figures (a)--(d) depict values for $K = 2$ to $5$ respectively. For even values of $K$, there must be at least one imaginary pair of solutions to preserve horizontal symmetry in the solutions. The number of solutions is equal to $2K-2$.}\label{fig:chid}
\end{center}
\end{figure}

In Figure \ref{fig:chid}, we present the values of $\chi'$ that solve the singulant equation for $K = 2$, $3$, $4$, and $5$. In cases where $K$ is even, the horizontal symmetry of solutions to \eqref{e:singulantK} ensures that there are at least two purely imaginary values of $\chi'$.

\begin{figure}
	\begin{center}
	\subfloat[$K=2$]{
\includegraphics{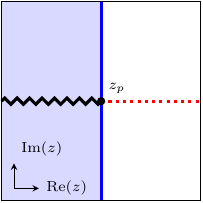}
	}
	\subfloat[$K=3$]{
\includegraphics{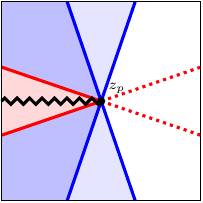}
	}
	\subfloat[$K=4$]{
\includegraphics{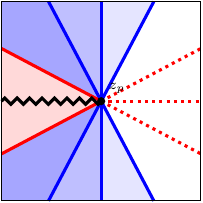}
	}
	\subfloat[$K=5$]{
\includegraphics{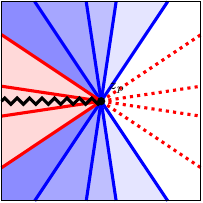}
	}
	\caption{Stokes structure around $z = z_p$ for different values of $K$. The legend is the same as in Figure \ref{fig:SL3}. The dark blue lines are Stokes lines, and the red lines are anti Stokes lines. As each Stokes line is crossed, a new exponentially small contribution appears. Darker blue shading indicates the presence of more exponentially small terms. As $K$ increases, the structure becomes more complicated as there are more Stokes lines and anti-Stokes lines originating at $z = z_p$. The red-shaded region, in which the power series expansion is invalid, grows as $K$ increases.}\label{fig:stokesK}
	\end{center}
	\end{figure}

In Figure \ref{fig:stokesK}, we show the Stokes structure around $z = z_p$ associated with the solutions for $\chi'$ for each value of $K$. The structures for $K=2$ and $K=3$ are shown in more detail in Figures \ref{fig:SL2} and \ref{fig:SL3}. From this figure, we see that increasing the family parameter $K$ increases the number of exponentials present in the problem, due to the number of solutions to the singulant equation (which is a polynomial of order $2K-2$) increasing at each order. Each solution for $\chi'$ produces more Stokes lines, leading to a more complicated Stokes structure as $K$ increases.

\section{Discrete Finite-Difference Equation}\label{s:FDinf}

We now calculate the Stokes switching that is present in the full finite-difference expression \eqref{e:FDfam}, corresponding to the $K\to\infty$ limit of \eqref{e:familyFD}. 

\subsection{Series Terms}
Substituting the power series expansion \eqref{e:dPIseries} into the infinite-order differential equation \eqref{e:FDfam} gives
\begin{align}
\sum_{m=1}^{\infty} \sum_{j=0}^{\infty} \frac{2\epsilon^{2m+2j-2}}{(2m)!}\diff{^{2m} y_j}{z^{2m}} + 3 \sum_{j=0}^{\infty}\sum_{l=0}^{\infty} \epsilon^{2j+2l}y_j y_l   = -2 z,
\end{align}
where $y_j$ is a function of $z$ for all $j$. By balancing at leading order in the limit that $\eps \to 0$, we find that $y_0$ satisfies the scaled Painlev\'{e} I equation, \eqref{e:scaleddP1}. As in the previous section, we select the tritronqu\'{e}e solution as the leading-order behaviour. Continuing to balance powers of $\eps$ at $\mathcal{O}(\eps^{2k})$ for $k > 0$ gives
\begin{align}\label{e:recurinf}
\sum_{m=1}^{k+1}\frac{2}{(2m)!}\diff{^{2m-2} y_{k-m+1}}{z^{2m-2}} + 3\sum_{l=0}^{k} y_l y_{k-l} = 0.
\end{align}
Unlike \eqref{e:recur}, the upper bounds of the summation terms always depends on $k$. 

\subsection{Late-Order Terms}

Finding the form of the late-order terms is more complicated than in the finite-$K$ case, as upper bound of the the summation terms in \eqref{e:recurinf} depends on $k$. 

\subsubsection{Singulant}

Substituting the late-order term ansatz \eqref{e:ansatz} into the recurrence relation \eqref{e:recurinf} and balancing at $\mathcal{O}(y_{k+1})$ as $k \to \infty$ gives
\begin{align}\label{e:singinf}
\sum_{m=1}^{k}\frac{2(\chi')^{2m-2}}{(2m)!} = 0.
\end{align}
In applying the late-order term ansatz, we are considering the asymptotic behaviour of the series terms as $k \to \infty$. This allows us to follow \cite{king_chapman_2001} and extend the range of summation to infinity. We write \eqref{e:singinf} as
\begin{align}\label{e:singinffull}
\cosh(\chi') - \sum_{m=k+1}^{\infty}\frac{2(\chi')^{2m-2}}{(2m)!}= 1.
\end{align}
Noting that $\chi'$ will be independent of $m$, it is clear that the second term is negligible as $k \to \infty$. Hence, the singulant equation becomes
\begin{align}\label{e:singinfext}
\sum_{m=1}^{\infty}\frac{2(\chi')^{2m}}{(2m)!} = 0.
\end{align}
Evaluating this sum gives
\begin{equation}\label{e:FDsing}
\cosh(\chi') = 1,
\end{equation}
and we recall that $\chi(z_p) = 0$. This has an infinite number of solutions, $\chi = 2M\pi\i(z-z_p)$ where $M \in \mathbb{Z}$. 

\begin{figure}
\begin{center}
\subfloat[Values of $\chi'$]{
\includegraphics{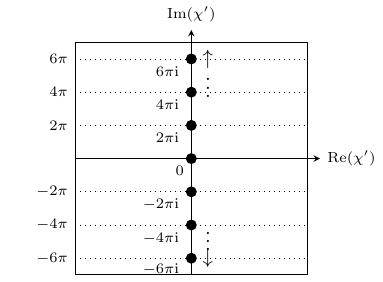}
}
\subfloat[Stokes structure near $z = z_p$]{
\includegraphics{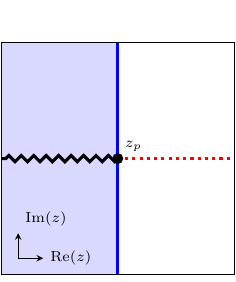}
}
\caption{Figure (a) depicts solutions of the singulant equation for $\chi'$ from the discrete finite-difference equation \eqref{e:FDfam}. There are solutions for all $\chi' = 2M\pi\i$ where $M \in\mathbb{Z}$. Figure (b) illustrates the corresponding Stokes structure near $z = z_p$, where the legend is the same as Figure \ref{fig:SL3}. There are two Stokes lines originating at $z=z_p$, one vertically upwards and one vertically downwards. An infinite number of distinct exponentially small contributions, one associated with each  $\chi'$, appears as the Stokes lines are crossed.}\label{fig:Kinf}
\end{center}
\end{figure}

The plot of $\chi'$ and the corresponding Stokes structure is given in Figure \ref{fig:Kinf}. All values of $\chi'$ are imaginary, so the solution contains only vertical Stokes lines and horizontal anti-Stokes lines, as in the $K=2$ case shown in Figure \ref{fig:SL2}. By comparing this with Figures \ref{fig:SL3} and \ref{fig:stokesK}, we see that the Stokes structure has become less complicated than $K = 3$, $K=4$ or $K=5$. 

\subsubsection{Prefactor}

If we return to the expression obtained by substituting the late-order term ansatz \eqref{e:ansatz} into \eqref{e:recurinf} and balance at $\mathcal{O}(y_{k+1/2})$ as $k\to\infty$, we obtain
\begin{align}\label{e:prefinf1}
\sum_{m=1}^{k}\frac{2(\chi')^{2m-3}Y'}{(2m-1)!} = 0,
\end{align}
where $\chi' = 2M\pi\i$ for some $M\in \mathbb{Z}$. If we apply the same method as \cite{king_chapman_2001} and extend the upper bound of the sum to infinity, we find that the resultant expression is zero, irrespective of the value of $Y'$. This unusual behaviour was seen in \cite{king_chapman_2001}, and no new information can be obtained by balancing at this order, in contrast to the finite-$K$ case. 

Balancing the expression at $\mathcal{O}(y_k)$ as $k \to \infty$ gives
\begin{align}\label{e:prefinf2}
\sum_{m=1}^{k}\binom{m}{2}\frac{2(\chi')^{2m-4}Y''}{(2m)!} + 6 y_0 Y= 0.
\end{align}
By extending the upper bound of the sum to infinity, this reduces to
\begin{align}\label{e:prefinf3}
Y''(z) + 6 y_0(z) Y(z)= 0.
\end{align}
This equation is the same as that obtained by linearizing \eqref{e:P1m} about a solution $y_0(z)$. If $y_0(z)$ is itself not available in closed form, we cannot obtain a closed-form solution to this equation, though information about $Y(z)$ is in principle derivable from the integrability of \eqref{e:P1m}; we shall not pursue such matters here. Using the local asymptotic behaviour of $y_0$ near poles \eqref{e:P1m_a2}, we find that if there is a pole at $z = z_p$ the local solution behaviour satisfies 
\begin{equation}
Y''(z)\sim \frac{12 Y(z)}{(z-z_p)^2} \quad \mathrm{as} \quad z \to z_p.
\end{equation}
Solving this gives
\begin{equation}\label{e:localpref}
Y(z) \sim (z-z_p)^4 \Lambda_1 + \frac{\Lambda_2}{(z-z_p)^3},
\end{equation}
where $\Lambda_1$ and $\Lambda_2$ are constants that have yet to be determined, which are associated with the two linearly independent solutions of \eqref{e:prefinf3}. 

Following \cite{king_chapman_2001}, we identify that the dominant leading-order behaviour of the late-order terms ansatz \eqref{e:ansatz} is associated with the first term in \eqref{e:localpref}, and hence $\Lambda_1$. Near $z = z_p$, the late-order ansatz associated with this term takes the form
\begin{equation}
y_j \sim \frac{ \Lambda_{1,1} (z-z_p)^4\Gamma(2j +\gamma)}{[2 \pi\i  (z-z_p)]^{2j+\gamma}} + \frac{\Lambda_{1,2}(z-z_p)^4 \Gamma(2j +\gamma)}{[-2 \pi\i  (z-z_p)]^{2j+\gamma}} \quad \mathrm{as} \quad j \to \infty, \, z \to z_p,
\end{equation}
where the dominant behaviour comes from the singulants with $M = \pm1$, which we will denote as $\chi_1$ and $\chi_2$. The corresponding constants are denoted $\Lambda_{1,1}$ and $\Lambda_{1,2}$. 

For this expression to be consistent with the leading-order behaviour $y_0$, we require that the strength of the singularity at $z = z_p$ be equal to 2 when $j$ is set to 0. This gives the result that $\gamma = 6$. This distinguishes the discrete problem from the finite-$K$ members of the finite-difference family, for which $\gamma = 2$. 

A similar analysis can be applied to the second term in \eqref{e:localpref}, associated with $\Lambda_2$, to give $\gamma = -1$. This is smaller than the value of $\gamma$ for $\Lambda_1$, Hence the factorial-over-power contribution \eqref{e:ansatz} associated with $\Lambda_2$  must be small in the limit $j \to \infty$ compared to the contribution associated with $\Lambda_1$.

\subsubsection{The constant $\Lambda_1$}
\begin{figure}[tbp]
\centering
\includegraphics{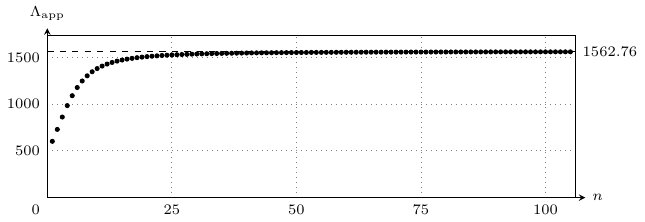}
\caption{Approximation of $\Lambda_1$ obtained by evaluating \eqref{e:recurKmatch} at different values of $k$, denoted $\Lambda_{\mathrm{app}}$. The value of $\Lambda_{\mathrm{app}}$ obtained as $k$ becomes large is shown as a dashed line.}\label{fig:Kinf2}
\end{figure}

{To calculate the constant $\Lambda_1$, we apply the same inner scaling as we did for the finite-$K$ problem, using the inner variables from \eqref{e:innervar}. This gives the discrete scaling for the finite-difference problem \eqref{e:FD} in terms of the scaled variable $v(\eta)$, and we find that the inner solution satisfies the leading-order equation
\begin{equation}\label{e:discretescale}
v(\eta+1) -2 v(\eta) + v(\eta-1) + 3 v(\eta)^2 = 0. 
\end{equation}}
The recurrence relation is found by expanding \eqref{e:discretescale} in the limit that $\eta \to \infty$ using \eqref{e:innerserK} and balancing powers of $\eta$. This gives 
\begin{equation}\label{e:recurKinf}
2\sum_{m=1}^{k+1}\binom{2k+3}{2m}v_{k-m+1} + 3 \sum_{\ell = 0}^{k}v_{\ell}v_{k-\ell} = 0.
\end{equation}
This expression is almost identical to \eqref{e:finiteKrecur}, except that the first summation continues to include new terms indefinitely as $k$ increases. We perform the matching in similar fashion, noting that
\begin{equation}\label{e:prefactormatchKinf}
\frac{v_{k}}{(z - z_p)^{{2k}+2}} \sim  \frac{ (\Lambda_{1,1}+\Lambda_{1,2})(z-z_p)^4\Gamma(2k+6)}{[2\pi \i (z - z_p)]^{2k+6}} \quad \mathrm{as} \quad k \to \infty.
\end{equation}
As there are no terms with odd singularity strength, we again conclude that $\Lambda_{1,1} = \Lambda_{1,2}$; we will subsequently denote both values as $\Lambda_1$. The matching condition \eqref{e:prefactormatchKinf} therefore gives
\begin{equation}\label{e:recurKmatch}
2\Lambda_1 = \lim_{k \to \infty} \frac{v_{k}(2\pi \i)^{2k+6}}{\Gamma(2k+6)}.
\end{equation}
We again approximate $\Lambda_1$ by calculating $v_{k}$ up to some large value of $k$, and evaluating the right-hand side of this expression. In Figure \ref{fig:Kinf2}, we illustrate the value of $\Lambda$ obtained by evaluating this expression for various $k$, denoted $\Lambda_{\mathrm{app}}$, illustrating the convergence of the approximation to a constant for larger values of $k$. Evaluating this expression at $k = 2000$ gives $\Lambda \approx 1562.76$. Hence, 
\begin{equation}\label{e:LOTKinf}
    y_k \sim \frac{\Lambda_1 \Gamma(2k+6)}{[2\pi\i(z - z_p)]^{2k+6}} + \frac{\Lambda_1 \Gamma(2k+6)}{[-2\pi\i(z - z_p)]^{2k+6}}\quad \mathrm{as}\quad k \to \infty, \qquad \Lambda_1 \approx 1562.76.
\end{equation}

\subsubsection{Stokes' switching}\label{S:discreteswitch}

To determine the quantity that is switched on across Stokes lines, we follow the method of \cite{king_chapman_2001}. Note that this is slightly more complicated than a typical Stokes' switching analysis. The details are given in detail in Appendix \ref{s:SwitchFD}. This analysis shows that the asymptotic quantity which is switched on across the vertical Stokes line extending from $z = z_p$, which satisfies $\mathrm{Im}(\chi) = 0$, and $\mathrm{Re}(\chi) > 0$ is given by
\begin{equation}
   R_N \sim \frac{Y\pi\i }{\epsilon^{\gamma}}  \e^{-\chi/\epsilon}.
\end{equation}
This is the exponentially small behaviour that is present in blue regions of Figure \ref{fig:Kinf}.

We note that, unlike the equivalent term for finite $K$ \eqref{e:RNapp}, the prefactor is not a constant. Instead, it is given by $Y(z)$, which in turn depends on the leading-order solution $y_0(z)$.

\subsection{Singulants for large finite $K$}

The increasing complexity of the Stokes structure as $K$ increases seen in Figure \ref{fig:stokesK} appears to be inconsistent with the simple Stokes structure seen in Figure \ref{fig:Kinf} (also seen in \cite{joshi2019generalized,moston2023nanoptera,king_chapman_2001} for example), in which the solutions for $\chi'$ are evenly spaced along the imaginary axis and all Stokes lines are vertical. 

\begin{figure}[tbp]
\centering
\subfloat[$K=10$]{
\includegraphics{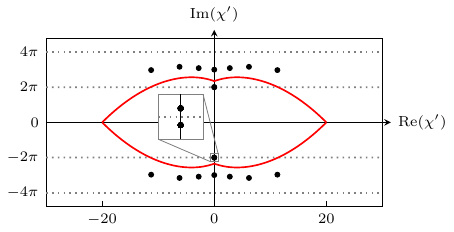}
 }
\subfloat[$K=25$]{
\includegraphics{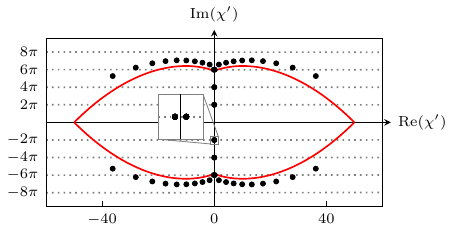}
 }
 
 \subfloat[$K=50$]{
\includegraphics{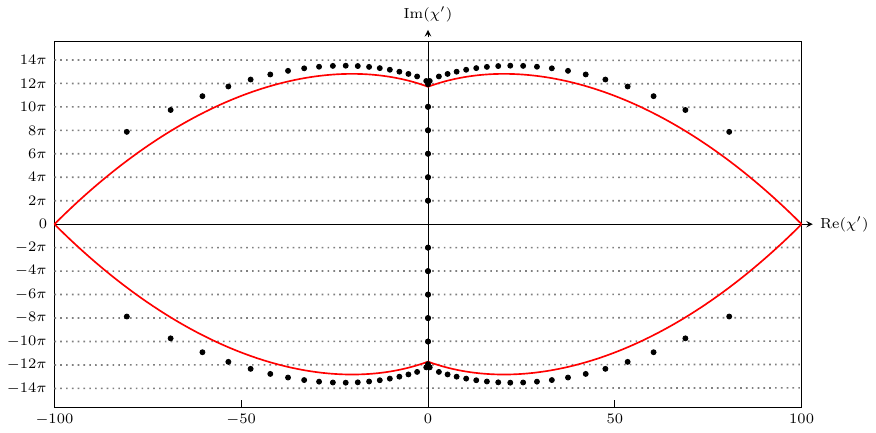}
 }
 
\caption{Solutions for $\chi'$ if (a) $K=10$, (b) $K=25$, and (c) $K=50$. Solutions are represented by filled circles. The red curves show the curve \eqref{e:Kbound}. Any values of $\chi'$ that do not approach the imaginary axis near some point $2\pi i n$ for some integer $n$ as $K$ goes large must be outside this curve. In (a) and (b) the inset shows a close-up diagram near $-2\pi\i$ that reveals two distinct solutions near this point. For even $K$, as in (a), these lie above and below $-2\pi\i$. For odd $K$, as in (b), these lie to the left and right of $-2\pi\i$. The solutions pairs get closer to to $2\pi\i n$ for fixed integer $n$ as $K$ increases. In (a) the width of the inset is roughly $10^{-1}$. In (b) the width of the inset is roughly $10^{-6}$.}\label{fig:largeK}
\end{figure}

In order to determine how the exponential behaviour for finite $K$ is consistent with the behaviour seen in the limit that $K \to \infty$, we solve the singulant equation \eqref{e:singulantK} and determine how the solutions $\chi'$ behave for large $K$. The results are shown in Figure \ref{fig:largeK} for $K = 10$, $K = 25$, and $K = 50$.

These figures indicate that as $K$ increases, there are values of $\chi'$ that tend to $\chi' = 2 \pi \i n$, for $n \in \mathbb{Z}$. This is consistent with the behaviour seen in Figure \ref{fig:Kinf} in the limit that $K \to \infty$. However, for sufficiently large $|\chi'|$, there are values of $\chi'$ that tend to a curve that does not follow the imaginary axis. These values are obtained by solving \eqref{e:singulantK}, {which is equivalent to finding zeros of the partial sum of $\cosh(z)$.}

{Approximating the zero locations of partial sums of $\e^z$ and related functions, such as $\cosh(z)$, is a classical problem first studied by Szeg\H{o} \cite{szego1924eigenschaft} and subsequently by others, including \cite{carpenter1991asymptotics,newman1972zeros,varga2010zeros} using direct analytic estimates, and more recently in \cite{kriecherbauer2008locating} using Riemann-Hilbert analysis. These studies show that the zero locations lie on rotated versions of the so-called Szeg\H{o} curve in the limit $K \to \infty$, or
\begin{equation}\label{e:Kbound}
|y| = \sqrt{4 K^2\, \e^{|x|/K-2}-x^2},
\end{equation}
where $\chi' = x + \i y$. This is equivalent to the asymptotic description of the zero locations of partial sums of $\cos(z)$ from \cite{varga2010zeros} rotated by $\pi/2$. The relevant Szeg\H{o} curves are labelled in red in Figures \ref{fig:largeK}, and scale proportionally to $K$ as $K\to \infty$. }

{While this is a previously known result, it is instructive to show that that can be obtained directly through the study of Stokes' phenomenon in the partial sums of $\cosh(z)$. In Appendix \ref{A.appking}, we show that zeros of the partial sums in \eqref{e:singulantK} that do not lie near the imaginary axis can be written in terms of solutions to a differential equation
\begin{equation}\label{eq:kingappODE}
\frac{1}{2K}\diff{\Phi(\zeta)}{\zeta} + \frac{1}{\zeta} = \Phi - 1.
\end{equation}
The solutions $\chi'$ map to values of $\zeta$ such that $\Phi(\zeta) = 0$. These solutions occur due to two different exponential contributions in the solution cancelling each other out, and therefore must be found near anti-Stokes lines in the solution to \eqref{eq:kingappODE} where the two contributions are the same size. Expressing these anti-Stokes lines in terms of $\chi' = x + \i y$ gives the result from \eqref{e:Kbound}. We note that the Stokes structure of the problem can be equivalently obtained through a steepest-descents analysis of integral expression from \cite{kriecherbauer2008locating} for the zeros of the partial sum of the exponential function, which was derived as the solution to a Riemann-Hilbert problem.}

For any fixed values of $x$, the value of $y$ grows without bound as $K \to \infty$, so these solutions for $\chi'$ will not occur in the limit that $K \to \infty$. Only the singularities on the imaginary axis remain, and we recover the Stokes structure shown in Figure \ref{fig:Kinf}. We therefore see that, although complicated Stokes structures are generated for finite values of $K$, the Stokes structure seen in discrete problems does emerge in the limit that $K \to \infty$.

\section{Discrete Painlev\'{e} I Family}\label{s:integrable}

In the previous section, we calculated the behaviour of Stokes phenomenon in a non-integrable discrete equation. In this section, we perform a similar analysis for the finite-$K$ family of equations associated with integrable discrete Painlev\'{e} I \eqref{e:family}, before studying the behaviour of the infinite-order differential equation \eqref{e:dP1t}. This will allow us to study the effect of integrability on Stokes phenomenon in the discrete limit $K \to \infty$.

\subsection{Finite $K$}

A significant portion of this analysis is nearly identical to the finite-difference family \eqref{e:familyFD}, and we will therefore outline only the key steps and identify any differences.

\subsubsection{Series terms}

We first substitute the power series expansion \eqref{e:dPIseries} into \eqref{e:family}, which gives 
\begin{align}
\sum_{m=1}^K \sum_{j=0}^{\infty} \frac{2\epsilon^{2m+2j-2}}{(2m)!}\diff{^{2m} y_j}{z^{2m}} + 3 \sum_{j=0}^{\infty}\sum_{l=0}^{\infty} \epsilon^{2j+2l}y_j y_l + 2 \sum_{j = 0}^{\infty} y_{j}\sum_{m=1}^K \sum_{l=0}^{\infty} \frac{\epsilon^{2m+2j+2l}}{(2m)!}\diff{^{2m}y_{l}}{z^{2m}}   = -2 z,\label{e:dpIsub}
\end{align}
where $y_j$ is a function of $z$ for all $j$. Balancing this expression at $\mathcal{O}(1)$ as $\epsilon \to 0$, we again obtain the leading order equation \eqref{e:P1m}, and again choose the tritronqu\'{e}e solution described in \eqref{e:P1m_a1}--\eqref{e:P1m_a2} as $y_0$. Balancing at $\mathcal{O}(\epsilon^{2k})$ with $k > K$ gives the recurrence relation
\begin{align}\label{e:recurdpI}
\sum_{m=1}^K\frac{2}{(2m)!}\diff{^{2m-2} y_{k-m+1}}{z^{2m-2}} + 3\sum_{l=0}^{k} y_l y_{k-l}  + 2 \sum_{j=0}^{\infty}\sum_{m=1}^{K-1}\frac{2 y_j}{(2m)!}\diff{^{2m}y_{k-m-j}}{z^{2m}} = 0. 
\end{align}

\subsubsection{Late-Order Terms and Stokes Structure}
We begin by substituting the late-order ansatz \eqref{e:ansatz} into \eqref{e:recurdpI}. The largest two orders in the limit that $k \to \infty$ are both identical to the finite-difference family, and therefore \eqref{e:laterecur} is also valid for this family of differential equations. By balancing as $k \to \infty$, we find that the singulant is given by solutions to \eqref{e:singulantK}, so the Stokes structure is identical to that presented in Figure \ref{fig:SL2} for $K=2$, Figure \ref{fig:SL3} for $K = 3$, and \eqref{fig:stokesK} for larger values of $K$.

Continuing to balance terms in \eqref{e:laterecur} as $k \to \infty$ shows that the prefactor satisfies $Y'(z) = 0$, and the consistency condition requires that $\gamma = 2$. To proceed, we must calculate $\Lambda$. If we apply the scaling \eqref{e:innervar} and the inner expansion \eqref{e:innerserK} with $v_0 = -2$, we obtain the equation
\begin{align}\nonumber
2\sum_{m=1}^{K}\frac{1}{(2m)!}\sum_{j=0}^{\infty}\frac{(2j+2m+1)!}{(2j+1)!}&\frac{v_j}{\eta^{2j+2m+2}} + 3\sum_{\ell = 0}^{\infty}\sum_{j=0}^{\infty}\frac{v_{\ell}v_j}{\eta^{2j + 2\ell + 4}} \\&+ 2\sum_{\ell=0}^{\infty}\frac{v_{\ell}}{\eta^{2\ell+2}}\sum_{m=1}^{K-1}\frac{1}{(2m)!}\sum_{j=0}^{\infty}\frac{(2j+2m+1)!}{(2j+1)!}\frac{v_j}{\eta^{2j+2m+2}} =0.
\end{align}
Note that this is different to the corresponding inner problem for the finite-difference family \eqref{e:innerFD}, as it includes contributions from the second summation term. By balancing this expression at {$\mathcal{O}(\eta^{-2k-4})$}, we obtain
\begin{equation}\label{e:dPfiniteKrecur}
2\sum_{m=1}^{\mathrm{min}(K,k+1)}\binom{2k+3}{2m}v_{k-m+1} + 3 \sum_{\ell = 0}^{k}v_{\ell}v_{k-\ell} + 2\sum_{\ell=0}^{k-1} \sum_{m=1}^{\mathrm{min}(K-1,k-\ell)}\binom{2k-2\ell+1}{2m}v_{\ell}v_{k-\ell-m}= 0.
\end{equation}
This gives the recurrence relation for $v_k$
\begin{align}\nonumber
2(k+3)(2k-1)\,v_k = 2\sum_{m=2}^{\mathrm{min}(K,k+1)}\binom{2k+3}{2m}&v_{k-m+1} + 3\sum_{\ell=1}^{k-1}v_{\ell}v_{k-\ell}\\&+ 2\sum_{\ell=0}^{k-1} \sum_{m=1}^{\mathrm{min}(K-1,k-\ell)}\binom{2k-2\ell+1}{2m}v_{\ell}v_{k-\ell-m}.
\end{align}

\begin{figure}[tbp]
\centering
\includegraphics{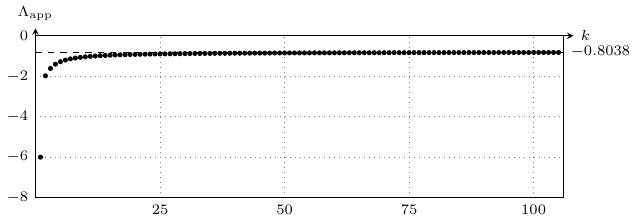}
\caption{Approximation of $\Lambda$ obtained by evaluating \eqref{e:K2match} at different values of $k$, denoted $\Lambda_{\mathrm{app}}$. The value of $\Lambda_{\mathrm{app}}$ obtained as $k$ becomes large is shown as a dashed line.}\label{fig:DPK2}
\end{figure}

For $K=2$, we can apply the matching condition \eqref{e:K2match} to approximate the constant $\Lambda$. In Figure \ref{fig:DPK2}, we illustrate the value of $\Lambda$ obtained by evaluating this expression for various $k$, denoted $\Lambda_{\mathrm{app}}$, illustrating the convergence of the approximation to a constant for larger values of $k$. Evaluating this expression at $k = 2000$ gives $\Lambda \approx -0.8038$.

Using an identical analysis to Section \ref{s:FDstokes}, we find that the exponentially small quantity that is switched on across the Stokes lines, $R_N$, satisfies the equation in \eqref{e:K3Rode}. Hence the jump across Stokes lines described in \eqref{e:finiteKjump} applies to the discrete Painlev\'{e} I family of equations. The only difference between the small exponential terms that appear in the finite-difference and discrete Painlev\'{e} I families is the value of $\Lambda$ that multiplies the exponential terms. 

In regions of the plane where the anti-Stokes lines are crossed and the exponential terms become large, corresponding to the red-shaded regions in Figure \ref{fig:SL3}, the additional terms present in the discrete Painlev\'{e} I family \eqref{e:family} have a more pronounced effect on the solution behaviour. A numerical comparison of the resulting behaviour is beyond the scope of the main text, but is presented in Appendix \ref{s:AppTrans} for $K=3$ and Appendix \ref{s:AppTransGen} for some larger values of $K$. This analysis provides numerical evidence that the interactions between growing exponential contributions produce moveable branch points to appear in the red-shaded regions. Moveable branch points are expected in solutions to \eqref{e:family} with finite $K$, as the differential equation is not integrable. In the integrable case, we will find that the solution does not contain Stokes' phenomenon, and therefore the growing exponentials, and hence moveable branch cuts, do not appear.

\subsection{Discrete Painlev\'{e} I Equation}

We now consider the discrete Painlev\'{e} I equation \eqref{e:dP1}. Prior asymptotic studies such as \cite{joshi1997local} have shown that equation has solutions which possess pole-free regions in the limit $n \to \infty$, similar to the tritronqu\'{e}e solutions to the continuous Painlev\'{e} equation. We therefore expect to find such regions when analysing the continuum limit equation \eqref{e:dP1t}. The asymptotic analysis of \eqref{e:dP1t} is nearly identical to that of the finite-difference equation \eqref{e:FDfam}, presented in Section \ref{s:FDinf}. 

Note that \cite{joshi1997local} shows that other leading-order solutions to \eqref{e:dP1} can produce pole-free regions; studying these would first require setting $w_n = (-1)^n v_n$ and studying the asymptotic behaviour of $v_n$ using the methods from the present study. This would be similar to the exponential asymptotic analysis from \cite{joshi2015, joshi2017} for a different discretised version of Painlev\'{e} I and Painlev\'{e} II.

In this analysis, we will show that the asymptotic solutions to the continuum limit equation \eqref{e:dP1t} do not contain any moveable singular points, unlike those seen in Figure \ref{fig:Kinf} (b), as the Stokes switching contribution that would cause such singularities to appear vanishes for discrete Painlev\'{e} I.

\subsubsection{Series terms}

We first substitute the power series expansion \eqref{e:dPIseries} into \eqref{e:family}, which gives 
\begin{align}
\sum_{m=1}^{\infty} \sum_{j=0}^{\infty} \frac{2\epsilon^{2m+2j-2}}{(2m)!}\diff{^{2m} y_j}{z^{2m}} + 3 \sum_{j=0}^{\infty}\sum_{l=0}^{\infty} \epsilon^{2j+2l}y_j y_l + 2 \sum_{j = 0}^{\infty} y_{j}\sum_{m=1}^{\infty} \sum_{l=0}^{\infty} \frac{\epsilon^{2m+2j+2l}}{(2m)!}\diff{^{2m}y_{l}}{z^{2m}}   = -2 z,\label{e:dpIinfsub}
\end{align}
where $y_j$ is a function of $z$ for all $j$. Balancing this expression at leading order as $\epsilon \to 0$ gives the scaled continuous Painlev\'{e} I equation \eqref{e:P1m}, and we again select the tritronqu\'{e}e solution described in \eqref{e:P1m_a1}--\eqref{e:P1m_a2} as $y_0$. Balancing at $\mathcal{O}(\epsilon^{2k})$ with $k > K$ gives the recurrence relation
\begin{align}\label{e:recurdpI2}
\sum_{m=1}^k\frac{2}{(2m)!}\diff{^{2m-2} y_{k-m+1}}{z^{2m-2}} + 3\sum_{l=0}^{k} y_l y_{k-l}  + 2 \sum_{j=0}^{\infty}\sum_{m=1}^{k-1}\frac{2 y_j}{(2m)!}\diff{^{2m}y_{k-m-j}}{z^{2m}} = 0. 
\end{align}

\subsubsection{Late-Order Terms and Stokes Structure}

The infinite-order equation \eqref{e:dP1t} is singularly perturbed as $\epsilon \to 0$, and the leading-order solution $y_0$ contains singularities. Normally, this means that the solution is described by a divergent asymptotic series due to the repeated differentiation of the series terms. If we assume that the series is indeed divergent, then we can apply the factorial-over-power ansatz presented in \eqref{e:ansatz} to determine their late-order behaviour.

We substitute the ansatz into \eqref{e:recurdpI2}. The largest two orders in the limit that $k \to \infty$ are both identical to those of the finite-difference equation \eqref{e:FDfam} from Section \ref{s:FDinf}. By balancing as $k \to \infty$, we find that the singulant satisfies the truncated series expression \eqref{e:singinf}. We extend the summation range to infinity (introducing only exponentially small error in the singulant) to obtain the singulant equation \eqref{e:FDsing}, which gives $\chi' = 2N\pi\i$ for $N \in \mathbb{Z}$. {This is the same as the equivalent result for the finite-difference problem, obtained by solving \eqref{e:FDsing}}, which implies that the Stokes structure is identical that presented in Figure \ref{fig:Kinf}. Continuing this analysis {in} identical fashion to Section \ref{s:FDinf} shows that the prefactor satisfies \eqref{e:prefinf3} and is described asymptotically by \eqref{e:localpref}. 

If the solution $y(z)$ has the Stokes structure depicted in Figure \ref{fig:Kinf} then it cannot satisfy the Painlev\'{e} property, typically associated with integrable ordinary differential equations. {Poles in the leading-order solution $z = z_p$ with divergent local expansions that generate Stokes' phenomenon correspond to essential singularities in the true solution \cite{Dingle}. Since the positions of these leading-order poles (and hence of essential singularities in the true solution) depend on the boundary conditions chosen for the differential equation, the solution contains moveable singularities that are not poles and therefore does not possess the Painlev\'{e} property.}

\subsubsection{Inner Problem}

{To determine the constant $\Lambda_1$ in the prefactor, we study the inner problem in the neighbourhood of the singularity at $z_p$. We express the problem in terms of the inner variables \eqref{e:innervar}, which again corresponds to the discrete scaling in the problem. This returns us to the discrete expression \eqref{e:dP1} in terms of the scaled inner variable $v(\eta)$, which has leading-order behaviour satisfying
\begin{equation}\label{e:discretescale2}
(1 + v(\eta))\left(v(\eta+1) + v(\eta) + v(\eta-1)\right) - 3 v(\eta) = 0.
\end{equation}
Equation \eqref{e:discretescale2} can be reduced to a first order difference equation (see Appendix \ref{S.A2Delta}), reflecting its status as integrable. We expand the solution as $\eta \to \infty$ to obtain the series for $y(\eta)$ in \eqref{e:innerserK}. Substituting this into \eqref{e:discretescale2} and balancing powers of $\eta$ gives 
\begin{equation}\label{e:dPinfiniteKrecur}
2\sum_{m=1}^{j+1}\binom{2k+3}{2m}v_{k-m+1} + 3 \sum_{\ell = 0}^{k}v_{\ell}v_{k-\ell} + 2\sum_{\ell=0}^{k-1} \sum_{m=1}^{k-\ell}\binom{2k-2\ell+1}{2m}v_{\ell}v_{k-\ell-m}= 0.
\end{equation}
 This gives the recurrence relation for $v_k$
\begin{align}\label{e:DeltaRecur}
-2(k+3)(2k-1)\,v_k = 2\sum_{m=2}^{k+1}\binom{2k+3}{2m}v_{k-m+1} + 3\sum_{\ell=1}^{k-1}v_{\ell}v_{k-\ell}+ 2\sum_{\ell=0}^{k-1} \sum_{m=1}^{k-\ell}\binom{2k-2\ell+1}{2m}v_{\ell}v_{k-\ell-m}.
\end{align}
In the finite-difference analysis, we used the recurrence relation \eqref{e:recurKinf} to approximate the value of $\Lambda_1$ using the matching condition \eqref{e:recurKmatch} by evaluating $v_k$ for large values of $k$. In the discrete Painlev\'{e} I equation, however, we can calculate $v_k$ exactly. It can be shown by direct substitution that
\begin{equation}
v_k = -2^{1-2k}.
\end{equation}
Note that $v_k$ decays as $k \to \infty$. In fact, we find that the series \eqref{e:innerserK} can be evaluated exactly, to give
\begin{equation}\label{e:D0series}
\sum_{j=0}^{\infty} \frac{v_j}{\eta^{2j+2}} = \frac{8}{4\eta^2-1}.
\end{equation}
It can be verified by direct substitution that this solves the difference equation \eqref{e:discretescale2}. If we attempt to use the matching condition \eqref{e:recurKmatch} to find $\Lambda_1$, we obtain
\begin{equation}
2\Lambda_1 = \lim_{k \to \infty} \frac{v_{k}(2\pi \i)^{2k+6}}{\Gamma(2k+6)} = 0.
\end{equation}
In this case, $\Lambda_1 = 0$, as $v_k$ decays as $k \to \infty$. Furthermore, as the inner expansion can be evaluated exactly, the resultant singularity is not an essential singularity, but instead is a pole. Hence, this behaviour does not violate the Painlev\'{e} property that all moveable singularities are poles.

This observation connects the discrete integrability of \eqref{e:dP1} with the requirement that integrable differential equations such as \eqref{e:dP1t} should satisfy the Painlev\'{e} property.

\section{General \texorpdfstring{$\Delta$}{Delta}}\label{s:GenDelta}
We recall that the more general family of differential equations \eqref{e:familygen} corresponds to the finite-difference family if $\Delta = -2$, and to the discrete Painlev\'{e} I family if $\Delta = 0$. In order to understand how the values of $\Lambda$ and $\Lambda_1$ change with $\Delta$, we study the values of the Stokes constant $\Lambda$ that appear as $\Delta$ is varied in the discrete problem (i.e. in the $K = \infty$ case).

Expressing the coordinates in inner variables \eqref{e:innervar} again returns us to the discrete expression \eqref{e:Delta1} in terms of the scaled inner variable $v(\eta)$, which has leading-order behaviour satisfying
\begin{equation}\label{e:discretescaleDel}
\left(1 + (1+\tfrac{\Delta}{2})v(\eta)\right)\left(v(\eta+1) -2 v(\eta) + v(\eta-1)\right) + 3 v(\eta)^2 = 0.
\end{equation}

For general $\Delta$, the series terms for \eqref{e:DeltaRecur} are generated using the recurrence relation
\begin{align}\nonumber
-2(k+3)(2k-1)\,v_k = 2\sum_{m=2}^{k+1}&\binom{2k+3}{2m}v_{k-m+1} + 3\sum_{\ell=1}^{k-1}v_{\ell}v_{k-\ell}\\
&+ (2+\Delta)\sum_{\ell=0}^{k-1} \sum_{m=1}^{k-\ell}\binom{2k-2\ell+1}{2m}v_{\ell}v_{k-\ell-m}.\label{e:recurDelta}
\end{align}
We can then calculate large values of $v_k$ and determine $\Lambda_1$ using the matching condition \eqref{e:K2match}. These values are shown in Figure \ref{fig:Delta}.
\begin{figure}[tbp]
\centering
\includegraphics[width=0.8\textwidth]{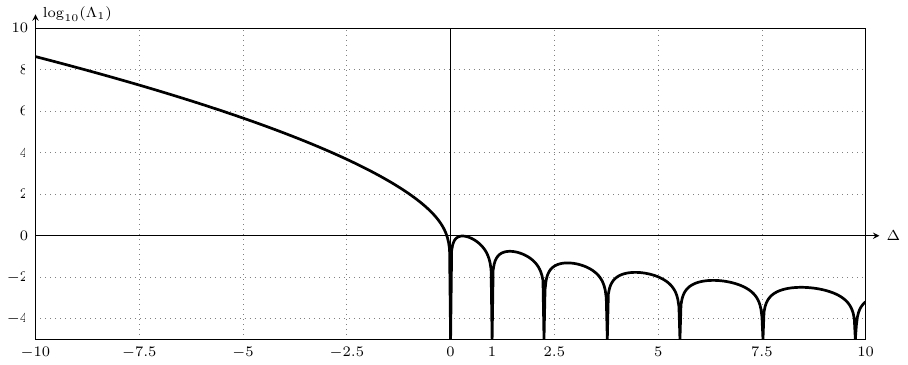}
\caption{Value of $\Lambda_1$ for a range of $\Delta$, shown on a logarithmic scale. $\Lambda_1$ is predicted using the matching condition \eqref{e:recurKinf} and values of $v_k$ computed from the recurrence relation \eqref{e:recurDelta}. This figure includes results for the finite-difference equation \eqref{e:FDfam}, $\Delta = -2$, and the discrete Painlev\'{e} I equation \eqref{e:dP1t},  $\Delta = 0$.}\label{fig:Delta}
\end{figure}
This figure predicts a number of isolated values of $\Delta$ that result in $\Lambda_1 = 0$. One of these values occurs for $\Delta = 1$, for which the recurrence relation for $v_k$ can be solved exactly to give $v_k = -2$. The series can be evaluated exactly, giving
\begin{equation}\label{e:D1series}
\sum_{j=0}^{\infty} \frac{v_j}{\eta^{2j+2}} = \frac{2}{\eta^2-1},
\end{equation}
which can be shown to solve \eqref{e:discretescaleDel} with $\Delta = 1$ by direct substitution. Note that while the coefficients $v_j$ do not decay as $j \to \infty$, they also do not grow factorially. Hence, the matching condition \eqref{e:K2match} still gives $\Lambda_1=0.$ 

In Appendix \ref{S.A2Delta}, we show that it is possible to reduce the $\Delta = 1$ inner equation \eqref{e:discretescaleDel} to a first-order difference equation, and that solutions to this difference equation can be generated by appropriately mapping solutions to the $\Delta = 0$ inner equation \eqref{e:discretescale2}. This suggests that the $\Delta = 1$ case also possesses integrable behaviour. Note that there are additional values of $\Delta > 1$ in Figure \ref{fig:Delta} that correspond to $\Lambda_1 = 0$. 

We note that in discrete problems such as these, there is a countably infinite number of terms that can be turned via the Stokes phenomenon (these arising from the Fourier modes associated with the lattice spacing, for which care is needed in adopting the infinite-order ODE representation). We conjecture that for the cases in Figure \ref{fig:Delta} that for which $\Lambda_1 = 0$ with $\Delta > 1$, higher modes \textit{will} be subject to Stokes switching, with the difference equation being non-integrable\footnote{This is similar to the behaviour of asymptotic solutions at transparent points seen in \cite{Alfimov}, where the first Fourier mode disappears, but higher modes exhibit Stokes switching}. That none of the modes will be switched on in the integrable case $\Delta = 0$ emphasises its very exceptional status; particularly given the results of Appendix \ref{S.A2Delta}, the case $\Delta = 1$ would seem to warrant further investigation.

Finally, we consider the asymptotic behaviour of the solution for values of $\Delta$ near zero. We linearize about the solution for $\Delta = 0$, such that 
\begin{equation}
v_k = -2^{1-2k} + \Delta \hat{v}_k, \qquad \hat{v}_k = 0.
\end{equation}
Retaining only the linear terms in $\Delta$ gives
\begin{align}\nonumber
 -2(k+3)(2k-1)\,\hat{v}_k =  2\sum_{m=2}^{k+1}\binom{2k+3}{2m}\hat{v}_{k-m+1}- 3\sum_{\ell=1}^{k-1}2^{2-2\ell}\hat{v}_{k-\ell} -  \sum_{\ell=0}^{k-1} \sum_{m=1}^{k-\ell}\binom{2k-2\ell+1}{2m}2^{2-2\ell}\hat{v}_{k-\ell-m}&\\
 -  \sum_{\ell=0}^{k-1} \sum_{m=1}^{k-\ell}\binom{2k-2\ell+1}{2m}2^{2-2(k-\ell-m)}\hat{v}_{\ell} + \sum_{\ell=0}^{k-1} \sum_{m=1}^{k-\ell}\binom{2k-2\ell+1}{2m}2^{2-2k+2m}&.
\end{align}
Calculating the terms $\hat{v}_k$ numerically provides evidence that $|\hat{v}_k|$ grows rapidly (see Figure \ref{fig:vk}). Hence, for any nonzero value of $\Delta$, the series terms eventually grow, causing the series coefficients $v_k$ to diverge factorially, and producing a nonzero value for $\Lambda_1$. This generates Stokes' phenomenon in the solution $y(z)$.

\begin{figure}[tbp]
\centering
\includegraphics{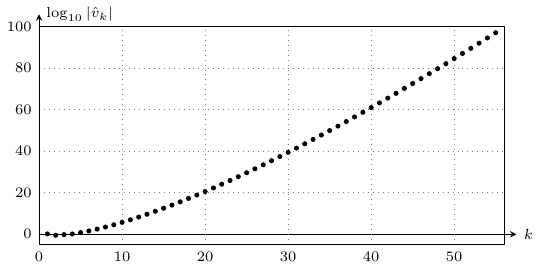}
\caption{Exponential growth of $|\hat{v}_k|$ as $k$ increases, shown on a logarithmic plot.}\label{fig:vk}
\end{figure}

\section{Conclusions}\label{s:conclusions}

This study used exponential asymptotics to resolve two questions about Stokes' phenomenon. The first question was how Stokes' phenomenon in discrete systems emerged from the solution of continuous systems as the order tends to infinity. The second question was how Stokes' phenomenon differs between integrable and non-integrable discrete systems, even when they tend to the same continuum limit.

In the first section of this paper, we studied Stokes' phenomenon present in a family of equations obtained by applying finite-difference discretisation to the Painlev\'{e} I equation, expanding this as a power series in the discretisation parameter, and truncating after a fixed number of terms, $K$. Our analysis showed that increasing $K$ caused Stokes' phenomenon in the solution to become more complicated, due to the increasing number of complex-valued solutions for the quantity $\chi'$. 

We then calculated Stokes' phenomenon present in the solution to the full, non-truncated equation, which only has imaginary-valued solutions for $\chi'$, producing a much simpler Stokes structure. By studying the behaviour of solutions for $\chi'$ as $K$ increases, we showed that as $K \to \infty$, the solutions for $\chi'$ either tend to the imaginary axis, or vanish. This explains the emergence of the simple Stokes structure in discrete problems that was observed in \cite{joshi2019generalized,moston2023nanoptera}.

In the second section of this paper, we performed a similar analysis for the continuum limit of discrete Painlev\'{e} I. We found that for finite $K$, Stokes' phenemonon present in the truncated differential equation family was nearly identical to that seen in the finite-difference equations, with the exponentially small terms differing only by a nonzero constant factor. 

When we repeated the exponential asymptotic analysis for the infinite-order differential equation generated from discrete Painlev\'{e} I, we determined that, unlike the finite-difference case, the asymptotic series for the solution does not diverge. In fact, the local solution near each pole is convergent, meaning that Stokes' phenomenon is not present in the solution. Hence, no subdominant exponentials appear; in the analysis of the finite-$K$ family, we showed numerical evidence that the exponentials would grow across anti-Stokes lines and generate moveable singular points in the asymptotic solution. The absence of Stokes' phenomenon in the solution to the the infinite-order differential equation means that these moveable singular points instead do not appear. This demonstrates that the asymptotic behaviour of the integrable discrete system in the continuum limit is consistent with continuous markers of integrability; in this case, the asymptotic solution does not contain moveable singular points. 

{The analysis in Section \ref{s:GenDelta} predicted that Stokes' phenomenon is not present in the $\Delta = 1$ case, among others. It should be noted that the $\Delta = 1$ case is different to the $\Delta = 0$ case in that the coefficients of the inner problem, $v_k$, do not decay as $k \to \infty$. For $\Delta = 0$ the coefficients were found to be $v_k = -2^{1-2k}$, while for $\Delta = 1$ the coefficients were $v_k = -2$. This means that the late-order terms do not vanish, but neither do they demonstrate factorial over power divergence. We did not pursue this question in the present study; however, it would be instructive to determine whether difference equations associated with $\Delta = 1$, such as \eqref{e:Delta1} are also integrable. }

{In each problem, determining the Stokes switching behaviour required us to rescale the problem to study the local behaviour of the solution near singular points. It is notable that for both the finite-difference and discrete Painlev\'{e} I equation, this inner problem recovered the leading-order discrete equation (see \eqref{e:discretescale} and \eqref{e:discretescale2}). In discrete problems, we typically find that $\chi'$ is a constant, and therefore that $\chi$ is a linear function of $z$. If $\chi \propto (z-z_s)$, the asymptotic series \eqref{e:dPIseries} must cease to be asymptotic if $z - z_s = \mathcal{O}(\eps)$. From this, we conclude that the inner scaling should be $\eta = \eps z$, which is precisely the original discrete scaling. We therefore expect that in discrete problems, Stokes switching in the corresponding infinite-order differential equation (eg. \eqref{e:dP1t}) is connected to the large-$|n|$ asymptotic behaviour of the original discrete equation (eg. \eqref{e:dP1}).}

{Noting this connection, it would be worthwhile in future studies to investigate whether there is a direct connection between discrete markers of integrability, such as singularity confinement, and continuous markers of integrability, such as the Painlev\'{e} property, in systems obtained by taking continuum limits of discrete integrable equations.}

\section{Acknowledgements}
CJL gratefully acknowledges ARC Discovery Project DP240101666. JRK gratefully acknowledges a Royal Society Leverhulme Trust Senior Fellowship. CJL and JRK would like to thank the Isaac Newton Institute for Mathematical Sciences, Cambridge, for support and hospitality during the programme ‘Applicable resurgent asymptotics: towards a universal theory’, where much of the work on this paper was undertaken. The programme was supported by the EPSRC grant no. EP/R014604/1. We are grateful to the referees for helpful comments, including regarding \cite{szego1924eigenschaft} and related work.

\appendix

\section{Stokes Switching Calculations}\label{s:AppSwitch}

In this section, we calculate the quantities that are switched on as Stokes lines are crossed in the complex plane, and therefore determine the asymptotic form of the exponentially small solution contribution in regions where the exponential is small. This analysis is identical for the finite-difference and discrete Painlev\'{e} I discretisations. Appendix \ref{s:SwitchFDK} gives the exponential terms for finite $K$, corresponding to the blue regions in Figures \ref{fig:SL2}, \ref{fig:SL3}, \ref{fig:stokesK}. {Appendix \ref{s:SwitchFD}} gives the form of the exponentially small asymptotic terms in the blue regions of Figure \ref{fig:Kinf}.

\subsection{Finite-Difference Family: Finite \texorpdfstring{$K$}{K}}\label{s:SwitchFDK}

Truncate the series \eqref{e:dPIseries} after $N$ terms, to obtain
\begin{equation}\label{e:truncatedser}
    y = \sum_{j = 0}^{N-1} \epsilon^{2j}y_j(z) + R_N(z).
\end{equation}
To determine the optimal truncation point, we follow the heuristic described in \cite{Boyd1999} and determine the value of $N$ such that $|\eps^{2N}y_N| \sim |\eps^{2N+2}y_{N+1}|$ for small $\epsilon$. We find that the optimal truncation is given by $N \sim |\chi|/2\epsilon$, so we set $N = |\chi|/2\epsilon + \omega$ where $\omega$ is bounded as $\epsilon \to 0$, and is chosen so that $N$ is an integer. Substituting the truncated series \eqref{e:truncatedser} back into the finite-difference equation \eqref{e:familyFD} gives, after some simplification
\begin{align}
    \sum_{m=1}^{K}\frac{2\epsilon^{2m-2}}{(2m)!} \diff{^{2m} R_N}{z^{2m}} & + 6 y_0 R_N + \ldots \sim - \frac{2\epsilon^{2N+2K-4}}{(2K)!}\diff{^{2K} y_{N-1}}{z^{2K}} + \ldots,
\end{align}
where the omitted terms are subdominant to those retained in the limit that $\epsilon \to 0$ and $N \to \infty$. We can apply the late-order ansatz and keep only the largest term on the right-hand side to obtain
\begin{align}\label{e:K3Rode}
   \sum_{m=1}^{K}\frac{2\epsilon^{2m-2}}{(2m)!} \diff{^{2m} R_N}{z^{2m}} & + 6 y_0 R_N \sim - \frac{2\epsilon^{2N+2K-4}}{(2K)!} \frac{(\chi')^{2K} \Lambda \Gamma(2N +2K)}{\chi^{2N + 2K}}.
\end{align}
This is a linear ordinary differential equation. The relveant complemetary function is given by
\begin{equation}
    R_N = C \Lambda \e^{-\chi/\epsilon},
\end{equation}
where $C$ an arbitrary constant, and the factor of $\Lambda$ is included for algebraic convenience. We will select the $\Lambda$ and $\chi$ associated with the Stokes line across which switching occurs. To take into account the right-hand side term, we permit the constant $C$ to vary (cf. reduction of order), and instead denote it by $\mathcal{S}(z)$, so that
\begin{equation}
    R_N = \mathcal{S}\Lambda \e^{-\chi/\epsilon}.
\end{equation}
Substituting this into \eqref{e:K3Rode} and simplifying using the singulant equation \eqref{e:singulant3} gives
\begin{equation}
-\left[\sum_{m=1}^{K}\frac{2(\chi')^{2m-1}}{(2m-1)!}\right]\frac{\Lambda}{\epsilon}\diff{\mathcal{S}}{z}\e^{-\chi/\epsilon} \sim - \frac{2\epsilon^{2N+2K-4}}{(2K)!} \frac{(\chi')^{2K} \Lambda \Gamma(2N +2K)}{\chi^{2N + 2K}}.  
\end{equation}
Note that we could simplify this further with known constant values for $\chi'$ and $\gamma$, but the analysis is clearer if we leave these in their current form. Changing the independent variable to $\chi$ and applying algebraic manipulation gives
\begin{equation}
    \diff{\mathcal{S}}{\chi} \sim  \frac{ \epsilon^{2N+2K-3} \e^{\chi/\epsilon} \Gamma(2N + 2K)}{\mu \chi^{2N + 2K}}, \quad \mathrm{where} \quad \mu = \sum_{m=1}^{K}\frac{2(2K)!}{(2m-1)!(\chi')^{2K-2m}}.
\end{equation}
Now we let $\chi = r \e^{\i\theta}$ and consider the variation only in $\theta$. Hence, we set
\begin{equation}
    \diff{}{\chi} = -\frac{\i \e^{-\i\theta}}{r}\diff{}{\theta}.
\end{equation}
We also use that the optimal truncation is given by $N = r/2\epsilon + \omega$. This gives
\begin{equation}
    \diff{\mathcal{S}}{\theta} \sim  \frac{\i r \epsilon^{r + 2\omega + 2K -3}\e^{\i\theta}}{\mu}  \frac{  \Gamma(r + 2\omega + 2K  )}{ (r\e^{\i\theta})^{r + 2\omega +2K}} \,\exp\left(\frac{r \e^{\i\theta}}{\epsilon}\right).
\end{equation}
Using Stirling's formula and simplifying this expression gives
\begin{equation}
    \diff{\mathcal{S}}{\theta} \sim \frac{\i\sqrt{2\pi r}}{\mu \epsilon^{3/2}} \exp\left(\frac{r}{\epsilon}(\e^{\i\theta}-1) - \frac{\i\theta r}{\epsilon} - \i\theta(2\omega + 2K +1)\right).
\end{equation}
We now let $\theta = \sqrt{\epsilon}\vartheta$, and find that
    \begin{equation}
    \diff{\mathcal{S}}{\vartheta} \sim \frac{\i\sqrt{2\pi r}}{\mu \epsilon^{2 }} \e^{-r \vartheta^2/2}.
\end{equation}
Integrating across the Stokes line gives
\begin{equation}\label{e:finiteKjump}
    [\mathcal{S}]_-^+ \sim \frac{2\pi\i}{\mu \epsilon^{2}}, \qquad [R_N]_-^+ \sim \frac{2\pi\i \Lambda}{\mu \epsilon^{2}} \e^{-\chi/\epsilon}.
\end{equation}

\subsection{Discrete Finite-Difference Equation}\label{s:SwitchFD}

Truncate the series \eqref{e:dPIseries} after $N$ terms, to obtain \eqref{e:truncatedser}. The optimal truncation is given by $N \sim |\chi|/2\epsilon$, so we set $N = |\chi|/2\epsilon + \omega$ where $\omega$ is bounded as $\epsilon \to 0$, and is chosen so that $N$ is an integer. Substituting the truncated series back into the finite-difference equation \eqref{e:FD} gives, after simplification,
\begin{align}
    \sum_{m=1}^{\infty}\frac{2\epsilon^{2m-2}}{(2m)!} \diff{^{2m} R_N}{z^{2m}} & + 6 y_0 R_N  \sim - 2 \sum_{p=N}^{\infty}\sum_{m=p-N+2}^{p+1}\frac{\epsilon^{2p}}{(2m)!} \diff{^{2m} y_{p-m+1}}{z^{2m}} + \ldots \quad \mathrm{as} \quad \eps \to 0,
\end{align}
where the omitted terms are subdominant to those retained in the limit that $\epsilon \to 0$ and $N \to \infty$. All remaining asymptotic relationships in this section shall be taken to be in the limit that $\epsilon \to 0$, which also corresponds to $N \to \infty$. 

From the range of the outer sum, we see that $p \geq N$ and therefore $p \gg 1$. Hence, we can extend the inner summation to infinity while only incurring errors that are exponentially small compared to $R_N$.  We apply the late-order ansatz and retain only the largest terms on the right-hand side to obtain 
\begin{equation}
\sum_{m=1}^{\infty}\frac{2\epsilon^{2m-2}}{(2m)!} \diff{^{2m} R_N}{z^{2m}}  + 6 y_0 R_N  \sim - 2 \sum_{p=N}^{\infty}\sum_{m=p-N+2}^{\infty}\frac{\epsilon^{2p}(\chi')^{2m}Y\Gamma(2p+\gamma+2)}{(2m)!\chi^{2p+\gamma+2}}.
\end{equation}
Setting $p = N + q$, noting that $\chi' = 2\pi\i$, and switching the order of summation gives
\begin{equation}
\sum_{m=1}^{\infty}\frac{2\epsilon^{2m-2}}{(2m)!} \diff{^{2m} R_N}{z^{2m}}  + 6 y_0 R_N  \sim  2 \sum_{m=2}^{\infty}\sum_{q=0}^{m-2}\frac{\epsilon^{2N+2q}(2\pi)^{2m}Y\Gamma(2N+2q+\gamma+2)}{(2m)!\chi^{2N+2q+\gamma+2}}.
\end{equation}
We now set $\chi = r\e^{\i\theta}$, which gives $N = r/2\epsilon + \omega$. This gives
\begin{equation}
\sum_{m=1}^{\infty}\frac{2\epsilon^{2m-2}}{(2m)!} \diff{^{2m} R_N}{z^{2m}}  + 6 y_0 R_N  \sim  2 \sum_{m=2}^{\infty}\sum_{q=0}^{m-2}\frac{\epsilon^{r/\epsilon+2\omega+2q}(2\pi)^{2m}Y\Gamma(r/\epsilon+2\omega+2q+\gamma+2)}{(2m)!\chi^{r/\epsilon+2\omega+2q+\gamma+2}}.
\end{equation}
Using Stirling's approximation, we may expand the gamma function to obtain
\begin{equation}
\sum_{m=1}^{\infty}\frac{2\epsilon^{2m-2}}{(2m)!} \diff{^{2m} R_N}{z^{2m}}  + 6 y_0 R_N  \sim  2 \sum_{m=2}^{\infty}\sum_{q=0}^{m-2}\frac{(2\pi)^{2m}Y\sqrt{2\pi}}{(2m)!\epsilon^{\gamma+3/2}r^{1/2}}\exp\left(-\frac{r}{\epsilon}-\i\theta\left(\frac{r}{\epsilon} + 2\omega+2q + \gamma+2\right)\right).
\end{equation}
The summation terms on the right-hand side can be evaluated exactly, giving
\begin{equation}\label{e:discreteStokes1}
\sum_{m=1}^{\infty}\frac{2\epsilon^{2m-2}}{(2m)!} \diff{^{2m} R_N}{z^{2m}}  + 6 y_0 R_N  \sim  \frac{2\sqrt{2\pi}Y}{\epsilon^{\gamma+3/2}r^{1/2}}\exp\left(-\frac{r}{\epsilon}-\i\theta\left(\frac{r}{\epsilon} + 2\omega+ \gamma\right)\right)g(\theta),
\end{equation}
where
\begin{equation}\label{e:gasymp}
g(\theta) = \frac{\cos\left(2\pi\e^{-\i\theta}\right)-1}{1-\e^{-2\i\theta}} \sim -\i\pi^2\theta \quad \mathrm{as} \quad \theta \to 0.
\end{equation}
The asymptotic behaviour of $g(\theta)$ as $\theta \to 0$ will be significant at a later stage in the analysis. We again set
\begin{equation}
    R_N = \mathcal{S}Y \e^{-\chi/\epsilon},
\end{equation}
where $\mathcal{S}$ varies rapidly in the vicinity of a Stokes line, while $Y$ varies slowly. The expression \eqref{e:discreteStokes1} becomes
\begin{align}
\frac{1}{\epsilon^2}\sum_{m=1}^{\infty}\left[\frac{2(\chi')^{2m}}{(2m)!} \mathcal{S}Y + \frac{2\epsilon(\chi')^{2m-1}}{(2m-1)!} \diff{(\mathcal{S}Y)}{z}+ \frac{\epsilon^2(\chi')^{2m-2}}{(2m-2)!} \diff{^2 (\mathcal{S}Y)}{z^2}\right]\e^{-\chi/\eps}+6 y_0 \mathcal{S}Y\e^{-\chi/\eps} &
\\
 \sim  \frac{2\sqrt{2\pi}Y}{\epsilon^{\gamma+3/2}r^{1/2}}\exp\left(-\frac{r}{\epsilon}-\i\theta\left(\frac{r}{\epsilon} + 2\omega+ \gamma\right)\right)&g(\theta).
\end{align}
Evaluating the sum and using \eqref{e:FDsing} and \eqref{e:prefinf3} to simplify the expression significantly gives 
\begin{align}
\diff{^2 (\mathcal{S}Y)}{z^2}\sim  \frac{2\sqrt{2\pi}Y}{\epsilon^{\gamma+3/2}r^{1/2}}\exp\left(\frac{r\e^{\i\theta}}{\epsilon}-\frac{r}{\epsilon}-\i\theta\left(\frac{r}{\epsilon} + 2\omega+ \gamma\right)\right)&g(\theta).
\end{align}
We follow similar steps to the analysis in Section \ref{s:SwitchFDK} and write the derivatives in terms of the variable $\theta$. The right-hand side is exponentially small except near $\theta = 0$, which we know to be the Stokes line. We also note that $Y$ varies slowly in the neighbourhood of the Stokes line, and may therefore be treated as approximately constant in $\theta$. This leaves
\begin{align}
\e^{-\i\theta}\diff{}{\theta}\left(\e^{-\i\theta} \diff{\mathcal{S}}{\theta}\right) \sim  \frac{2\sqrt{2\pi}Y}{\epsilon^{\gamma+3/2}r^{1/2}}\exp\left(\frac{r\e^{\i\theta}}{\epsilon}-\frac{r}{\epsilon}-\i\theta\left(\frac{r}{\epsilon} + 2\omega+ \gamma\right)\right)&g(\theta).
\end{align}
To capture the variation in the neighbourhood of the Stokes line, we again set $\theta = \sqrt{\epsilon}\vartheta$, which gives after some rearrangement
\begin{equation}
\frac{1}{\epsilon}\diff{^2\mathcal{S}}{\vartheta^2} \sim -\frac{\i r^{3/2}}{\epsilon^{\gamma+1}}\sqrt{\frac{\pi}{2}}\vartheta \e^{-r\vartheta^2}{2},
\end{equation}
where the asymptotic behaviour of $g(\theta)$ as $\theta \to 0$ from \eqref{e:gasymp} was used to simplify the right-hand side. Integrating this expression twice gives 
\begin{equation}\label{e:StokesKinf}
    [\mathcal{S}]_-^+ \sim \frac{\pi\i}{\epsilon^{\gamma}}, \qquad [R_N]_-^+ \sim \frac{Y\pi\i }{\epsilon^{\gamma}}  \e^{-\chi/\epsilon}.
\end{equation}
Since $R_N = 0$ on the right-hand side of the Stokes lines in Figure \ref{fig:Kinf}, we know that the exponentially small contribution on the left-hand side is given by $R_N = [\mathcal{S}]_-^+$ from \eqref{e:StokesKinf}.

\section{Approximating solutions to the singulant equation for \texorpdfstring{$\chi'$}{χ'}}\label{A.appking}

We wish to approximate solutions to the partial sums of $\cosh(\chi') - 1$, or solutions to
\begin{equation}\label{eq:john0}
\sum_{n=1}^K \frac{2 (\chi')^{2n-2}}{(2n)!} = 0,
\end{equation}
in the asymptotic limit $K \to \infty$. To locate these zeroes, we will first consider the partial sums of the exponential function,
\begin{equation}\label{eq:john1}
\phi^+(z) = \sum_{n=0}^N\frac{z^n}{n!},
\end{equation}
in the limit that $N \to \infty$. We can rewrite this expression as
\begin{equation}\label{eq:john1.5}
\phi^+(z) =  \e^z  - \sum_{n=N+1}^{\infty}\frac{z^n}{n!}.
\end{equation}

{Locating zeroes of the partial sums of $\e^z$ is a classical problem was first studied by Szeg\H{o} \cite{szego1924eigenschaft} and later by others \cite{carpenter1991asymptotics,newman1972zeros,kriecherbauer2008locating}; this can be directly extended to locate zeroes of related functions such as $\cos(z)$ \cite{varga2010zeros}. Early direct analytic estimates showed that the solutions in the limit $K \to \infty$ lie either on the imaginary axis or on curves in the complex plane known as Szeg\H{o} curves.}

{The integral analysis in \cite{kriecherbauer2008locating} reveals that the zeros of the partial sums of the exponential function can be located by performing steepest-descents analysis on an integral expression, and therefore that they can be understood by studying the Stokes' structure of the solution. In this section, we present a derivation of this Stokes structure, and the associated zero locations, using differential equation methods; this approach can be conveniently generalised to other more complicated discrete singulant equations. Where appropriate, we will note the different regimes identified in \cite{kriecherbauer2008locating} and how they relate to the regimes found in the present analysis.}

\subsection{$z \in \mathbb{R}^+$}\label{s:kingreal}

{In this section, we first utilise results from the NIST Digital Library of Mathematical Functions \cite{NIST:DLMF} and the references therein, as well as contributions from \cite{nemes2019asymptotic}, which contains a comprehensive history of the relevant asymptotic approximations. We use 
\begin{equation}\label{eq:phiQ}
\phi^+(z) = \e^{z} Q(N+1,z),
\end{equation}
where $Q(a,z) = \Gamma(a,z)/\Gamma(a)$ \cite[\href{https://dlmf.nist.gov/8.4.E10}{(8.4.10)}]{NIST:DLMF}. We then set $z = N\zeta$ and make use of known expansions.  The asymptotic behaviour of $Q(a,b)$ as $a \to \infty$ is catalogued in \cite[\href{https://dlmf.nist.gov/8.12}{\S 8.12}]{NIST:DLMF}, and predicts three relevant regions to consider: in the rescaling, these are $0 < \zeta < 1$, $\zeta > 1$, and a transition region near $\zeta =1$. }

{In the first two regimes, the asymptotic behaviour of $\phi^+$ is given by {the expansion for $\Gamma(a) - \Gamma(a,z)$ in \cite[\href{https://dlmf.nist.gov/8.11.E6}{(8.11.6)}]{NIST:DLMF}} and the expansion for $\Gamma(a,z)$ in \cite[\href{https://dlmf.nist.gov/8.11.E7}{(8.11.7)}]{NIST:DLMF} respectively. This gives
\begin{align}\label{eq:newasymp}
\phi^+ \sim 
\begin{cases}
{\e^{N\zeta}} & \mathrm{if} \quad 0 < \zeta < 1,\\
\dfrac{\e^N \zeta^N}{\sqrt{2\pi N}}\dfrac{\zeta}{\zeta - 1} & \mathrm{if} \quad 1 < \zeta < \infty,
\end{cases} 
\end{align}
as $N \to \infty$. Note that the asymptotic behaviour of $\phi^+$ if $\zeta < 1$ is singular in the limit $\zeta \to 1$, therefore predicting a transition regime near this point.}

{The asymptotic behaviour in the transition regime is more complicated, and was rigorously established in \cite{nemes2019asymptotic}, in which the authors studied the asymptotic behaviour of $Q(a,z)$ in the transition regime. This work found that the transition regime occurs (in our notation) at $\zeta = 1 + N^{-1/2}\xi$ with $\xi = \mathcal{O}(1)$, and the asymptotic behaviour as $N \to \infty$, {catalogued in \cite[\href{https://dlmf.nist.gov/8.12.E4}{(8.12.4)}]{NIST:DLMF}}, is
\begin{equation}\label{eq:johnmiddle00}
\phi^+ \sim   \frac{1}{2}\mathrm{erfc}\left(\frac{\xi}{\sqrt{2}}\right){\e^{N+N^{1/2}\xi}} .
\end{equation}
This correctly matches to the first and second regime as $\xi \to -\infty$ and $\xi \to +\infty$, respectively. }

{Note that in \cite{kriecherbauer2008locating}, the integral expression for the partial sum $\phi^+$ predicts a solution that jumps on either side of a critical point at $z = 1$; this corresponds to the point $\zeta = 1$ in our notation. The location of this transition region is equivalent to the effect of the critical point in the Riemann-Hilbert integral formulation.}

{We next present an alternative derivation the above results in a way that we hope provides additional intuition as well as allowing for a self-contained analysis.}

\subsubsection{First Regime: $0 < \zeta < 1$} In this region, the behaviour of $\phi^+$ is dominated by a saddle point associated with $n = N\zeta + N^{1/2}\sigma$, where $\sigma = \mathcal{O}(1)$ as $N \to \infty$. The saddle point can be located by writing
\begin{equation}
\phi^+(\zeta) = \sum_{n=0}^N \e^{g(n)},\qquad g(n) = n \log(N \zeta) - \log(n!).
\end{equation}
and solving for $g'(n) = 0$. Expanding the solution for $N \to \infty$ gives the required result, including the value of $\sigma$, which is not significant to the present analysis. Since $N^{1/2} \gg 1$, we may apply a continuum limit approximation to \eqref{eq:john1.5} and integrate over $\sigma \in \mathbb{R}$. We know that the saddle point will lie within the range of integration, so this will be the dominant contribution to the asymptotic behaviour. We use the result 
\begin{equation}
g(n) \sim n \log(N) + n \log(\zeta) - n \log(n) + n \sim N \zeta - \frac{\sigma^2}{2\zeta}-\log(\sqrt{2\pi N \zeta})
\end{equation}
to obtain
\begin{equation}
\phi^+(\zeta) \sim \frac{1}{\sqrt{2\pi N\zeta}}\e^{N\zeta}\sum_{n=0}^N \e^{-\sigma^2/2\zeta}.
\end{equation}
We can then apply a continuum limit, and integrate over $\mathbb{R}$ to find, as $N \to \infty$,
\begin{equation}
\phi^+ \sim \frac{\e^{N\zeta}}{\sqrt{2\pi \zeta}}\int_{-\infty}^{\infty} \e^{-\sigma^2/2\zeta} \d \sigma = \e^{N \zeta} = \e^z.
\end{equation}

\subsubsection{Second Regime}

If $\zeta > 1$, the saddle point lies outside of the range of integration. In this case, the dominant asymptotic contribution must arise due to the end point of the sum. It is straightforward to see that it must come from the upper end point, as the summand increases until the saddle is reached. We shift the summation by writing it in terms of $m$, where $n = N-m$ and $m = \mathcal{O}(1)$.

We now have
\begin{equation}\label{eq:johnmsum}
\phi^+(z) = \sum_{m=0}^N \frac {z^{N-m}}{(N-m)!} = \frac{z^N}{N!}\sum_{m=0}^N  \frac{N!z^{-m}}{(N-m)!}.
\end{equation}
The dominant contribution now comes from $m \ll N$, so we can extend the summation to infinity without changing the asymptotic behaviour. We again use $z = N \zeta$ and find 
\begin{equation}
\frac{z^N}{N!}\sum_{m=0}^N  \frac{N!z^{-m}}{(N-m)!} \sim \frac{z^N}{N!}\sum_{m=0}^{\infty}\frac{N!z^{-m}}{(N-m)!} \sim  \frac{{(N\zeta)}^N}{N!}\sum_{m=0}^{\infty}\zeta^{-m}.
 \end{equation}
Evaluating the sum exactly gives
\begin{equation}\label{eq:regime2}
\phi^+(\zeta) \sim \frac{{(N\zeta)}^N}{N!}\frac{\zeta}{\zeta-1} \sim \frac{\e^N \zeta^N}{\sqrt{2\pi N}}\frac{\zeta}{\zeta - 1}.
\end{equation}
This has no poles in the region $\zeta > 1$, but does predict one at {$\zeta = 1$}, which indicates that the important intermediate region, corresponding to the critical point in \cite{kriecherbauer2008locating}, lies near this point.

\subsubsection{Distinguished Regime}

From the variation we see in the first regime, we know that the transition region between the two regimes must occur at $\zeta = 1 + N^{-1/2}\xi$ with $\xi = \mathcal{O}(1)$. This translates to $n = N - N^{1/2}\nu$ with $\nu = \mathcal{O}(1)$, or  $m = N^{1/2}\nu$. We will use this scaling so that we can again take advantage of the continuum limit. We have 
\begin{equation}
\phi^+(\xi) = \sum_{n=0}^N \e^{g(n)},\qquad g(n) = (N-m) \log(N + N^{1/2}\xi) - \log((N-m)!).
\end{equation}
Using a similar argument to before, we find that the saddle point occurs at $m = N^{1/2}\nu$, where $\nu = \mathcal{O}(1)$. While we can calculate the value of $\nu$, it is not needed here. Using similar reasoning, we are able to expand $g(n)$ in the limit that $N \to \infty$ to find that
\begin{equation}
g(n) \sim N + N^{1/2}\xi - \log(\sqrt{2\pi N})+ \frac{(\xi+\nu)^2}{2}.
\end{equation}
This allows us to write \eqref{eq:johnmsum} as an integral, and hence find as $N \to \infty$ that
\begin{equation}\label{eq:johnmiddle}
\phi^+ \sim  \frac{\e^{N + N^{1/2}\xi}}{\sqrt{2\pi}}\int_{0}^{\infty} \e^{-(\xi + \nu)^2/2\zeta} \d \sigma = \frac{1}{2}\mathrm{erfc}\left(\frac{\xi}{\sqrt{2}}\right){\e^{N+N^{1/2}\xi}}.
\end{equation}
This has the correct asymptotic behaviour to match to the first and second regime as $\xi \to -\infty$ and $\xi \to +\infty$, respectively as expected.

\subsection{$z \notin \mathbb{R}^+$}

{We now wish to consider values of $z$ that do not lie on the real axis. The key difference between this and the analysis above is that the saddle point no longer lies in the range of integration, and therefore analytic continuation is required to approximate the behaviour. In this section we again identify the critical point at $\zeta = 1$, and determine the associated Stokes structure, analytically continuing behaviour off the real axis using an ordinary differential equation approach. The use of a  differential equation approach is different to preceding analyses of this problem, such as \cite{carpenter1991asymptotics,newman1972zeros,kriecherbauer2008locating,szego1924eigenschaft}.}

\subsubsection{Governing Differential Equation}

Let us consider 
\begin{equation}\label{eq:john1.5.2}
\phi^+(z) =  \e^z  - \sum_{n=N+1}^{\infty}\frac{z^n}{n!}
\end{equation}
We begin by differentiating \eqref{eq:john1.5.2} with respect to $z$, to obtain
\begin{equation}
\diff{\phi^+}{z} = \e^z - \sum_{n=N+1}^{\infty} \frac{z^{n-1}}{(n-1)!} = \phi^+ - \frac{z^N}{N!}.
\end{equation}
As before, we work in terms of $\zeta$, where $z = N \zeta$. This will ensure that the distinguished region still corresponds to $\zeta$ near 1. We apply the change of variables $\phi^+(z) = z^N\Phi(\zeta)/N!$ to obtain
\begin{equation}
\frac{(N\zeta)^N}{(\zeta N)!}\Phi + \frac{z^N}{N!}\diff{\zeta}{z}\diff{\Phi}{\zeta} = \frac{z^N\Phi}{N!} - \frac{z^N}{N!},
\end{equation}
which simplifies to give
\begin{equation}\label{eq:johnODE}
\frac{1}{N}\diff{\Phi}{\zeta} + \frac{1}{\zeta}\Phi = \Phi - 1.
\end{equation}
Note that for $N = 0$ we obtain $\phi^+ = 1$. This gives the boundary condition $\Phi \to 1$ as $\zeta \to \infty$.

\subsubsection{Asymptotic Contributions}

We write the solution as
\begin{equation}
\Phi(\zeta) \sim \Phi_0(\zeta) + \frac{1}{N}\Phi_1(\zeta) + \ldots
\end{equation}
 as $N \to \infty$. The first few orders of the solution are given by
\begin{equation}\label{eq:johnPhi}
\Phi \sim \frac{\zeta}{\zeta-1} + \frac{1}{N}\frac{\zeta}{(\zeta-1)^3} \qquad \mathrm{as} \qquad N \to \infty.
\end{equation}
This is the same result to leading order as the analysis from the second regime, given in \eqref{eq:newasymp}. Furthermore, it is clear that the asymptotic analysis breaks down if $\zeta = 1 + \mathcal{O}(N^{-1/2})$, suggesting the appropriate distinguished scaling $\zeta = 1 + N^{-1/2}\xi$, where $\xi = \mathcal{O}(1)$, and $\Phi(\zeta) = \Psi(\xi)/N^{1/2}$. The rescaled differential equation becomes
\begin{equation}
\diff{\Psi}{\xi} = \frac{\xi}{1+N^{-1/2}\xi} \Psi - 1.
\end{equation}
The boundary condition is obtained by matching the behaviour in the limit that $\xi \to \infty$ with \eqref{eq:johnPhi} as $\zeta \to 1$. The solution to this equation has leading-order behaviour as $N \to \infty$, denoted $\Psi_0(\xi)$, that satisfies
\begin{equation}
\diff{\Psi_0}{\xi} = \xi \Psi_0 - 1,
\end{equation}
giving (cf. \eqref{eq:johnmiddle})
\begin{equation}
\Psi_0(\xi) = C \e^{\xi^2/2} + \sqrt{\frac{\pi}{2}}\e^{\xi^2/2} \mathrm{erfc}\left(\frac{\xi}{\sqrt{2}}\right),
\end{equation}
where $C$ is a parameter determined by the asymptotic matching condition, which gives $C = 0$. Hence,
\begin{equation}
\Psi_0(\xi) =\sqrt{\frac{\pi}{2}}\e^{\xi^2/2} \mathrm{erfc}\left(\frac{\xi}{\sqrt{2}}\right),
\end{equation}
or, in the original coordinates,
\begin{equation}\label{eq:johnmiddle0}
\Phi(\zeta) \sim \sqrt{\frac{\pi N}{2}}\e^{N(\zeta-1)^2/2} \mathrm{erfc}\left(\frac{N(\zeta-1)}{\sqrt{2}}\right).
\end{equation}
As $N \to \infty$ for $\zeta < 1$, the asymptotic behaviour of this expression is
\begin{equation}\label{eq:johnmiddle2}
\Phi(\zeta) \sim \sqrt{2\pi N}\e^{N(\zeta-1)^2/2}  \qquad \mathrm{as} \qquad N \to \infty.
\end{equation}
To confirm that this expression has the appropriate behaviour for $\zeta < 1$, we note that $\phi \sim \e^z$ as $N \to \infty$ in this region. This means that
\begin{equation}\label{eq:johnlead}
\Phi \sim \frac{N! \e^z}{z^N} = \frac{N! \e^{N\zeta}}{(N\zeta)^N} \sim \sqrt{2\pi N}\e^{N(\zeta -1 - \log(\zeta))}.
\end{equation}
As $\zeta \to 1$, this behaviour is consistent with the intermediate region \eqref{eq:johnmiddle}. 

For most of the complex plane, there are two significant asymptotic contributions to the solution, given in \eqref{eq:johnPhi} and \eqref{eq:johnlead}. We will subsequently denote these expressions as $\Phi^{(1)}$ and $\Phi^{(2)}$ respectively. The expression \eqref{eq:johnmiddle0} describing the behaviour in the neighbourhood of the turning point at $\zeta = 1$; we will denote this as $\Phi^{(3)}$. Note that neither $\Phi^{(1)}$ nor $\Phi^{(2)}$ predict zeroes in the solution, and any zeroes must arise due to interactions between these contributions. $\Phi^{(3)}$ does, however, have zeroes, a number of these being reported in \cite{NIST:DLMF}.

To determine the location of the zeroes we require, we need to study the Stokes switching behaviour that occurs between $\Phi^{(1)}$ and $\Phi^{(2)}$. We will briefly note the impact on the solution behaviour from $\Phi^{(3)}$ at the conclusion of the analysis.

\subsubsection{Stokes Structure}

From \eqref{eq:johnlead}, we see that the solution contains a Stokes line that satisfies
\begin{equation}
\mathrm{Im}(\zeta - 1 - \log(\zeta)) = 0, \qquad \mathrm{Re}(\zeta - 1- \log(\zeta)) > 0,
\end{equation}
and an anti-Stokes line that satisfies
\begin{equation}\label{eq.JohnAS}
\mathrm{Re}(\zeta - 1 - \log(\zeta)) = 0.
\end{equation}
These curves are presented in Figure \ref{fig:kingstokesfig}. {Note that this is precisely the condition that would be obtained using a steepest-descents approach to the the integral equation from \cite{kriecherbauer2008locating} to determine whether the saddle point contributes to the approximation, and confirming the methods agree. }

We see that the contribution $\Phi^{(2)}$ cannot be present on the right of the Stokes line, because it would dominate the behaviour on the real axis for $\zeta > 1$ if it were present. Hence, it appears as the Stokes line is crossed, and is asymptotically subdominant compared to the dominant contribution $\Phi^{(1)}$ when it appears.  Crossing the anti-Stokes line towards the origin causes the relative dominance of the two contributions to change, which means that $\Phi^{(2)}$ is the dominant contribution inside the anti-Stokes line, in the region that includes the origin and the real axis for $0 < \zeta < 1$. This This behaviour is consistent with the results we obtained for $z \in \mathbb{R}^+$ in Section \ref{s:kingreal}.

\begin{figure}[tbp]
\centering
\includegraphics[width=0.5\textwidth]{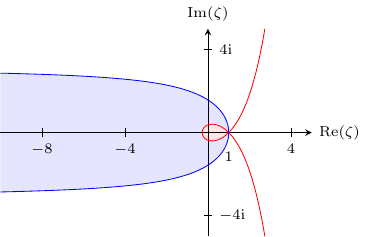}
\caption{Stokes structure of asymptotic solution to \eqref{eq:johnODE}. Stokes lines are shown in blue, and anti-Stokes lines are shown in red. In unshaded regions, the asymptotic solution contains only the contribution $\Phi^{(1)}$. In shaded regions, the asymptotic solution contains both $\Phi^{(1)}$ and $\Phi^{(2)}$. In the blue-shaded region, the dominant asymptotic contribution is $\Phi^{(1)}$, while in the red-shaded region, the dominant asymptotic contribution is $\Phi^{(2)}$.}\label{fig:kingstokesfig}
\end{figure}

\subsubsection{Zero Locations}

In the vicinity of the anti-Stokes lines, $\Phi^{(1)}$ and $\Phi^{(2)}$ are comparable in asymptotic size as $N \to \infty$. The asymptotic leading-order behaviour here is therefore given by $\Phi \sim \Phi^{(1)} + \Phi^{(2)}$. This means that zeroes of $\Phi$ can only exist near the anti-Stokes line \eqref{eq.JohnAS}, occurring due to cancellation between $\Phi^{(1)}$ and $\Phi^{(2)}$. 

If we let $\zeta= \zeta_r + \i \zeta_i$, we can express the anti-Stokes line as
\begin{equation}
\zeta_r - 1 - \tfrac{1}{2}\log(\zeta_r^2 + \zeta_i^2) = 0.
\end{equation}
Solving this and setting $\zeta_r = x/N$ and $\zeta_i = y/N$ gives
\begin{equation}\label{eq.johnAS1}
y = \pm\sqrt{N^2 \e^{2x/N-2} - x^2},
\end{equation}
where $z = x + \i y$. We will use this result to approximate the curves on which zeroes are found for the more complicated expression \eqref{eq:john0}.

To approximate the actual solutions to $\phi^+(z) = 0$ for large $|z|$, we set $\zeta = r\e^{\i\theta}$ and look for solutions to $\Phi^{(1)}(\zeta) = -\Phi^{(2)}(\zeta)$, or
\begin{equation}\label{e:johnsols}
\frac{r\e^{\i\theta}}{r\e^{\i\theta} - 1} = -\sqrt{2\pi N}\e^{N(r\exp(\i\theta) - 1 - \log(r) - \i\theta)}.
\end{equation}
Some algebra gives
\begin{equation}
r\exp(\i\theta) - 1 - \log(r) - \i\theta = \frac{1}{N}\log\left(\frac{1}{\sqrt{2\pi N}} \frac{\zeta}{\zeta - 1}\right) + \frac{2\pi \i M}{N},
\end{equation}
where $M \in \mathbb{Z}$. Further algebraic manipulation gives
\begin{equation}\label{eq:johnrtheta-1}
r\exp(\i\theta) - 1 - \log(r) - \i\theta =- \frac{\log(2\pi N)}{2N} + \frac{2\pi\i M }{2N} + \frac{1}{N}\log\left(\frac{\zeta}{\zeta - 1}\right).
\end{equation}
Let let $\log(\zeta/(\zeta-1)) = \log(\kappa) + \i\sigma$, where $\kappa \in \mathbb{R}$ and $\sigma \in (-\pi,\pi]$. Matching real and imaginary parts of this equation gives a system of equations for $\kappa$ and $\sigma$,
\begin{align}\label{eq:johnrtheta0}
r\cos(\theta) - \log(r) - 1 = - \frac{\log(2\pi N)}{2N}  + \frac{\log(\kappa)}{N}, \qquad r\sin(\theta) - \theta  = \frac{2\pi M }{2N}  + \frac{\sigma}{N},
\end{align}
and therefore that
\begin{align}\label{eq:johnrtheta1}
\kappa \cos(\sigma) + \frac{r^2 - r\cos(\theta)}{r^2 - 2r \cos(\theta)+1} =0, \qquad
\kappa \sin(\sigma) - \frac{r\sin(\theta)}{r^2 - 2r \cos(\theta)+1}.
\end{align}
We are interested in the case where $M = \mathcal{O}(N)$ as $N \to \infty$, which means that we can balance \eqref{eq:johnrtheta0} at leading-order to obtain the system
\begin{align}\label{eq:johnrtheta}
r \cos(\theta) - \log(r) - 1 = 0, \qquad r\sin(\theta) - \theta = \frac{2\pi M}{N},
\end{align}
which we can solve to determine the leading-order value of $\zeta$ along which the expressions are balanced, and hence approximate locations of solutions to $\phi^+(z) =0$. 

{It is useful to set $\theta = \pi$, for which \eqref{eq:johnrtheta} gives $M = -N/2$ and $r = \tau$ where $\tau$ solves $1 + \tau-\log\tau)$; this is the point where the anti-Stokes line in \eqref{fig:kingstokesfig} intersects with the real axis, so all zeroes must be to the right of this point. As we expect the zeroes to appear in complex conjugate pairs, this is consistent with the polynomial \eqref{eq:john1} having $N$ zeroes.}

{This choice of $\theta$ and $r$ may be substituted into \eqref{eq:johnrtheta1}, which gives $\sigma = \pi$ and $\kappa = \tau/(\tau+1)$. Substituting these results into \eqref{eq:johnrtheta0} gives $M = -(N+1)/2$. This means that a real zero is present in the solution only if $N$ is odd, whereas if $N$ is even then there are no zeroes on the real axis, and the zeroes appear solely in complex conjugate pairs. }

{By setting $\theta = \pi/2$, we follow a similar approach and show that $r = \e^{-1}$ and $M = -N/4 + Nr/(2\pi)$. This indicates that we expect to find $N/4 - N/(2\pi\e)$ zeroes in the first quadrant of the $\zeta$-plane, and hence $\lfloor N/2 - N/(\pi\e)\rfloor $ zeroes in the right half plane.}

{The zeroes of \eqref{e:johnsols} for $N = 40$ are presented in Figure \ref{fig:johnzeros}. This figure contains 40 zeroes, which appear in complex conjugate pairs follow the anti-Stokes line more closely on the left-hand side, and diverge from the curve as they approach the transition region near $z = 1$.  There are no real-valued zeroes, as $N$ is even. The previous analysis predicts that there are approximately $\lfloor 7.658 \rfloor$, or 7 zeroes in the first quadrant, which is equal to the value shown in the figure.}

{Note that the anti-Stokes lines in Figure \ref{fig:kingstokesfig2} reach the points $z = \pm N$, where we expect that the solution to instead be described by $\Phi^{(3)}$. This explains why the numerical solutions in Figure \ref{fig:johnzeros} diverge from the anti-Stokes line as these points are approached. The behaviour of the zeroes as the intermediate regime is approached can be calculated from $\Phi^{(3)}$ in \eqref{eq:johnmiddle}, which governs the solution near the turning point. Zeroes of $\Phi^{(3)}$ must be calculated numerically; however, numerical zeroes of the complementary error function nearest to the origin are documented, and can be found in \cite{NIST:DLMF}. }

\begin{figure}[tbp]
\centering
\includegraphics{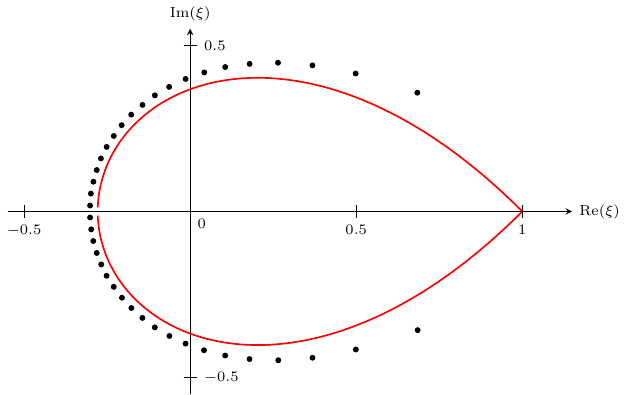}
\caption{Solutions to \eqref{e:johnsols} for $N=40$ are shown as black circles. The anti-Stokes line \eqref{eq.JohnAS} is shown as a red curve. The zeroes of \eqref{e:johnsols} follow the anti-Stokes curves more accurately as the move further from the transition region near $\zeta = 1$. There are 40 zeroes that appear in complex conjugate pairs, with 7 zeroes contained in the first quadrant.}
\label{fig:johnzeros}
\end{figure}

\subsection{Zeroes of singulant equation}

In addition to \eqref{eq:john1.5.2}, we define
\begin{equation}
\phi^-(z) = \phi^+(-z) = \e^{-z} - \sum_{n = N+1}^{\infty} \frac{(-z)^n}{n!}.
\end{equation}
This function has the same Stokes structure shown in \eqref{fig:kingstokesfig}, reflected across the vertical axis, using the mapping $z \mapsto -z$. Hence, the zeroes of $\phi^-$ must lie on the curve
\begin{equation}\label{eq.johnAS2}
y = \pm\sqrt{N^2 \e^{-2x/N-2} - x^2}.
\end{equation}
We now write
\begin{equation}
\psi(z) = \frac{1}{2}\left(\phi^+ + \phi^-\right) = \cosh(z) - \frac{1}{2}\sum_{n = N+1}^{\infty}\frac{z^{n}}{n!}- \frac{1}{2}\sum_{n = N+1}^{\infty}\frac{(-z)^{n}}{n!} = \sum_{n=0}^{N}\frac{z^{2n}}{(2n)!}.
\end{equation}

If $\mathrm{Re}(z) > 0$ and $|z| \gg 0$, then $\psi \sim \tfrac{1}{2}\phi^+$ as $N \to \infty$. The anti-Stokes lines in \eqref{eq.johnAS1} take are located at $|z| = \mathcal{O}(N)$ as $N \to \infty$, and are therefore present in the Stokes structure of $\psi(z)$ for $\mathrm{Re}(z) > 0$. This implies that solutions of $\phi^+(z) = 0$ for $\mathrm{Re}(z) > 0$ must also be approximate solutions to $\psi(z) = 0$. 

A similar argument can be made for $\mathrm{Re}(z)< 0$ and $|z| \gg 0$, where $\psi \sim \tfrac{1}{2}\phi^-$ as $N \to \infty$. In this region, the anti-Stokes lines of $\phi^-(z)$, and hence solutions of $\phi^-(z) = 0$ must also be approximate solutions to $\psi(z) = 0$. We therefore predict that solutions for $|z| \gg 0$ are found along the curves \eqref{eq.johnAS1} if $\mathrm{Re}(z) > 0$ and \eqref{eq.johnAS2} if $\mathrm{Re}(z) < 0$. Combining these relations shows that zeroes of $\psi(z)$ for large $|z|$ approximately lie on
\begin{equation}\label{eq.johnAS3}
y = \pm\sqrt{N^2 \e^{2|x|/N-2} - x^2}.
\end{equation}
This behaviour is shown in Figure \ref{fig:kingstokesfig2}. Note that neither approximation is valid if $|\mathrm{Im}(z)| = \mathcal{O}(1)$ as $N \to \infty$.

\begin{figure}[tbp]
\centering
\includegraphics{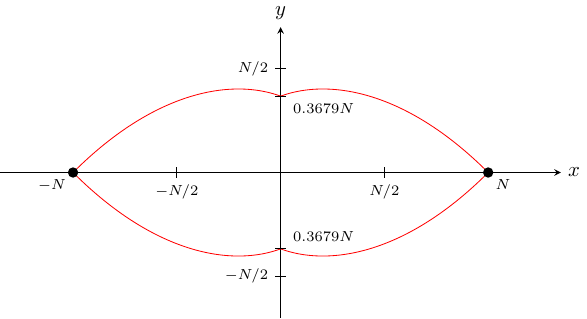}
\caption{Anti-Stokes lines of $\psi(z)$ {in the limit that} $N \to \infty$, where $z = x + \i y$. Solutions to \eqref{eq:john0} for large $|z|$ are expected to be located approximately on the anti-Stokes lines, shown in red. The Stokes lines always have a distance from the origin that is proportional to $N$.}\label{fig:kingstokesfig2}
\end{figure}

Replacing $N$ with $2K$ gives, after some algebra,
\begin{align}
\psi(z) = \sum_{n=0}^{N}\frac{z^{2n}}{(2n)!} = \frac{z^2}{2}\left(\sum_{n=1}^{K}\frac{2 z^{2n-2}}{(2n)!} + \frac{2}{z^2}\right).
\end{align}
For $z$ which is large in the limit that $K \to \infty$, the final term is $\mathcal{O}(K^{-2})$, and hence can be neglected. This means that solutions of $\psi(z)$ for large $|z|$ approximately satisfy
\begin{equation}
\sum_{n=1}^{K}\frac{2 z^{2n-2}}{(2n)!} = 0,
\end{equation}
which is precisely the expression from \eqref{eq:john0}. Hence, applying $N = 2K$ to \eqref{eq.johnAS3} shows that solutions to \eqref{eq:john0} for large $|z|$ must asymptotically approach the curve
\begin{equation}
y = \pm\sqrt{4K^2 \e^{|x|/K-2} - x^2}
\end{equation}
in the limit $K \to \infty$. 

{Finally, recall that the number of zeros $\phi^+(z)$ possesses $\sim N/2 - N/(\pi\e)$ in the right-half plane. Therefore $\psi(z)$ contains $\sim N - 2N/(\pi\e)$ zeroes away from the imaginary axis. Given that we expect the solution to contain $N$ {zeroes} in total, we expect that it will contain $\sim  2N/(\pi\e)$ zeroes below the anti-Stokes line near the imaginary axis. This is consistent with the expectation that the anti-Stokes lines intersect the imaginary axis at $\mathrm{Im}(z) = N/\e$ and that there are two zeroes in the vicinity of each multiple of $2\pi\i$, whereby $\cosh(z) \to 0$ as $N \to \infty$, within $|\mathrm{Im}(z)| < N/\e$.}

{These results could be obtained using the analytic estimates from \cite{varga2010zeros} or the steepest-descents analysis from \cite{kriecherbauer2008locating}. Nonetheless, we believe that it is instructive to see directly how the zeroes relate to the Stokes structure found from the differential equation solved by the partial sums. This method is straightforward to generalise to other partial sums that typically appear in singulant equations to discrete problems.}

\section{Transasymptotic Analysis for \texorpdfstring{$K=3$}{K=3}}\label{s:AppTrans}

\subsection{Finite-Difference Family}\label{S:AppFDK3}

We begin with the differential equation from \eqref{e:3}. The Stokes structure for this equation is presented in Figure \ref{fig:SL3}. The purpose of this analysis is to construct a transseries approximation for the behaviour in the blue-shaded region of Figure \ref{fig:SL3}(c). We will then use transasymptotic resummation to continue this approximation into the red-shaded region of the plane, and finally use Pad\'{e} approximation to provide numerical evidence that the red-shaded region contains moveable branch points.

\subsubsection{Transseries}

In the blue-shaded region of Figure \ref{fig:SL3} (c), the solution contains subdominant exponential contributions as $\eps \to 0$. We will study the exponentials that grow as we cross the anti-Stokes line in the positive angular direction from the real axis, crossing into the red-shaded region. We write the solution in the limit $\epsilon \to 0$ in the form of a transseries as
\begin{equation}\label{e:tseries}
y \sim \Phi_0(z; \epsilon) + \sum_{q=1}^{\infty}\epsilon^{\beta_q} \e^{-q \chi(z)/\epsilon}\Phi_q(z; \epsilon) \quad \mathrm{as} \quad \epsilon \to 0,
\end{equation}
where 
\begin{equation}
\Phi_q(z; \epsilon) \sim \sum_{m=0}^{\infty}\epsilon^{2m} \Phi_{q}^{(m)}(z),
\end{equation}
and $\Phi_{q}^{(m)} \neq 0$. This transseries captures not just the largest exponentially small term, but also the smaller exponentials associated with this value of $\chi$; in nonlinear problems, the transseries should contain exponential terms with exponents that are integer multiples of the original exponential term \cite{aniceto2019primer}. The term $\Phi_0(z; \epsilon)$ corresponds to the leading-order power series solution, and the exponentially small remainder term \eqref{e:finiteKjump} gives the leading-order behaviour of $\Phi_1(z; \epsilon)$. 

Recall that $\chi'(z)$ is constant; for convenience, we let $\chi' = \alpha$. The first anti-Stokes line we encounter if we move in the positive angular direction is the anti-Stokes line associated with $\alpha = \sqrt{-15 + 3\i\sqrt{15}}$, or $\chi_3 $ from \eqref{e:allsingulants}. 

We substitute the transseries \eqref{e:tseries} into the governing equation \eqref{e:3} to find
\begin{align}\nonumber
\frac{1}{360}\left[\epsilon^4 \diff{^6 \Phi_0}{z^6}+ \sum_{q=1}^{\infty}\epsilon^{\beta_q-2}(q\chi')^6\e^{-q\chi/\epsilon}\Phi_q + \ldots\right] +\frac{1}{12}\left[\epsilon^2 \diff{^4 \Phi_0}{z^4} + \sum_{q=1}^{\infty}\epsilon^{\beta_q-2}(q\chi')^4\e^{-q\chi/\epsilon}\Phi_q + \ldots\right]&  \\
+\left[\diff{^2 \Phi_0}{z^2} + \sum_{q=1}^{\infty}\epsilon^{\beta_q-2}(q\chi')^2\e^{-q\chi/\epsilon}\Phi_q + \ldots\right]  +\left[\Phi_0 + \sum_{q=1}^{\infty}\epsilon^{\beta_q}\e^{-q\chi/\epsilon}\Phi_q\right]\left[\Phi_0 + \sum_{q=1}^{\infty}\epsilon^{\beta_q}\e^{-q\chi/\epsilon}\Phi_q\right ] &= -2z,
\end{align}
where the terms denoted by the ellipsis in each set of braces are smaller than those retained in the limit $\eps \to 0$. By balancing the dominant behaviour of the linear terms and the nonlinear term at each exponential order, we obtain a relation for $\beta_q$, where
\begin{equation}
\beta_q - 2 = \beta_1 + {\beta_{q-1}}.
\end{equation}
We know from the form of the remainder term in \eqref{e:finiteKjump} that $\beta_1 = -2$.  This recurrence relation is therefore solved by $\beta_q = -2$ for $q \geq 1$, giving the algebraic scaling for each transseries term.

We next define a transseries parameter $\xi$, where
\begin{equation}\label{e:transpar}
\xi = \sigma \e^{-\chi/\epsilon},
\end{equation}
where $\sigma$ is a scaling constant to be determined. Using \eqref{e:transpar}, we can write \eqref{e:tseries} as
\begin{equation}\label{e:tseries2}
y \sim \frac{1}{\epsilon^2}\sum_{q=0}^{\infty}\xi^{q} \Phi_q(z; \epsilon) \quad \mathrm{as} \quad \epsilon \to 0.
\end{equation}
This has the effect of scaling each term by a factor of $\sigma^q$, but we can absorb this scaling into the $\Phi_q$ term.  In the unshaded regions in Figure \ref{fig:SL3}, there are no exponential terms; we therefore set $\sigma = 0$.

If we continue in the positive angular direction across the Stokes line into the blue-shaded region of Figure \ref{fig:SL3}, the remainder term \eqref{e:finiteKjump} appears, which is the leading-order asymptotic behaviour of the series with $q=1$. Matching this leading-order behaviour with the remainder term at leading-order gives
\begin{equation}
\frac{\xi}{\epsilon^2} {\Phi_1^{(0)}(z)} \sim \frac{2\pi\i\Lambda}{\mu\epsilon^2}\e^{-\chi/\epsilon} \quad \mathrm{as} \quad \epsilon \to 0,
\end{equation}
which, using \eqref{e:transpar}, gives
\begin{equation}
\sigma \Phi_1^{(0)}(z)  = \frac{2\pi\i\Lambda}{\mu}.
\end{equation}
We are free to pick $\sigma$, and this expression suggests choosing $\sigma = 2\pi\i \Lambda /\mu$, which gives $\Phi_1^{(0)}(z) = 1$. 

The transseries \eqref{e:tseries2} is valid in regions where the exponential term $\xi$ is asymptotically small as $\xi \to 0$ (or zero), corresponding to the unshaded and blue regions shown in Figure \ref{fig:SL3}(b). In order to determine the behaviour in the red region, we write the double expansion for $y$ in the blue region of Figure \ref{fig:SL3}(b),
\begin{equation}\label{e:tseries3}
y \sim \frac{1}{\epsilon^2}\sum_{q=0}^{\infty}\sum_{m=0}^{\infty}\xi^{q} \epsilon^{2m} \Phi_q^{(m)}(z) \quad \mathrm{as} \quad \epsilon \to 0, \, \xi \to 0,
\end{equation}
where we note that $\Phi_0^{(0)} = 0$ so that the $q=0$ series is $\mathcal{O}(1)$ in the limit that $\eps \to 0$. This is a valid asymptotic expansion in the region where $\xi$ is small in the limit $\epsilon \to 0$.

\subsubsection{Resummation}

Now that we have the transseries \eqref{e:tseries} in the region where $\xi$ is small in the limit $\epsilon \to 0$, we may reverse the order of summation to obtain
\begin{equation}\label{e:tseries4}
y \sim \frac{1}{\epsilon^2}\sum_{m=0}^{\infty}  \epsilon^{2m} \sum_{q=0}^{\infty}\xi^{q} \Phi_q^{(m)}(z) \quad \mathrm{as} \quad \epsilon \to 0, \, \xi \to 0.
\end{equation}
We may now evaluate the inner summation term, to obtain
\begin{equation}
y \sim \frac{1}{\epsilon^2}\sum_{m=0}^{\infty}  \epsilon^{2m} \Phi^{(m)}(z; \xi), \qquad \Phi^{(m)} = \sum_{q=0}^{\infty}\xi^{q} {\Phi}_q^{(m)}(z).
\end{equation}
Evaluating the inner sum permits us to continue the transseries expression analytically into regions where $\xi$ is not small in the limit that $\epsilon \to 0$. Note that this is an established procedure which can alternatively be replicated by using traditional matched asymptotic analysis to continue to the solution between regions as different contributions contribute to the dominant asymptotic behaviour. The outer sum remains asymptotically valid outside of regions where $\xi$ is small, and therefore allows us to determine the asymptotic solution behaviour as $\eps \to 0$ even after the anti-Stokes line is crossed. In this region of the plane, the solution behaviour is
\begin{equation}\label{e:summedseries}
y \sim \frac{1}{\epsilon^2}\sum_{m=0}^{\infty}  \epsilon^{2m} \Phi^{(m)}(z; \xi) \quad \mathrm{as} \quad \epsilon \to 0,
\end{equation}
with no asymptotic dependence on $\xi$. Now we can substitute \eqref{e:summedseries} into the governing equation and obtain an equation for $\Phi^{(0)}$ (as well as $\Phi^{(m)}$). This is the only term that generates new singularities, as equations for $\Phi^{(m)}$ are linear. 

Note that $z$ and $\xi$ are not independent, and we can use \eqref{e:transpar} to write all derivatives in terms of $\xi$. Applying the chain rule gives
\begin{equation}
\diff{f}{z} = \diff{\xi}{z} \diff{f}{\xi} = -\frac{\alpha\xi}{\epsilon} \diff{f}{\xi}.
\end{equation}
We can substitute \eqref{e:summedseries} into \eqref{e:3} and use this expression to write the resultant equation in terms of $\xi$. Balancing this expression at $\mathcal{O}(\epsilon^{-4})$ gives the equation for ${\Phi}^{(0)}$,
\begin{align}\nonumber
\frac{\alpha^6}{360}\xi^6\diff{^{6}\Phi^{(0)}}{\xi^{6}}& 
+ \frac{\alpha^{6}}{24}\xi^5\diff{^{5}\Phi^{(0)}}{\xi^{5}}
+ \left(\frac{13\alpha^6}{72}+\frac{\alpha^4}{12}\right)\xi^4\diff{^{4}\Phi^{(0)}}{\xi^{4}}
+ \left(\frac{\alpha^6}{4}+\frac{\alpha^4}{2}\right)\xi^3\diff{^{3}\Phi^{(0)}}{\xi^{3}}\nonumber\\
&+ \left(\frac{31\alpha^6}{360}+\frac{7\alpha^4}{12}+\alpha^2\right)\xi^2\diff{^{2}\Phi^{(0)}}{\xi^{2}}
+ \left(\frac{\alpha^6}{360}+\frac{\alpha^4}{12}+\alpha^2\right)\xi\diff{\Phi^{(0)}}{\xi}
+ 3\left(\Phi^{(0)}\right)^2 = 0,\label{e:Psi0}
\end{align}
The solution to this equation is the analytic continuation of the inner sum in \eqref{e:tseries4} outside of the asymptotic limit $\xi \to 0$. In Section \ref{S.C1.3}, we will use this to study the singular points that appear in regions where the exponential terms, and hence $\xi$, are not small. We will obtain all necessary boundary conditions by matching this expression to the known transseries in the region where $\xi$ is small.

Continuing to balance terms gives differential equations for the corrections. The next correction satisfies
\begin{align}\nonumber
\frac{\alpha^6}{360}\xi^6\diff{^{6}\Phi^{(1)}}{\xi^{6}}& 
+ \frac{\alpha^{6}}{24}\xi^5\diff{^{5}\Phi^{(1)}}{\xi^{5}}
+ \left(\frac{13\alpha^6}{72}+\frac{\alpha^4}{12}\right)\xi^4\diff{^{4}\Phi^{(1)}}{\xi^{4}}
+ \left(\frac{\alpha^6}{4}+\frac{\alpha^4}{2}\right)\xi^3\diff{^{3}\Phi^{(1)}}{\xi^{3}}\nonumber\\
&+ \left(\frac{31\alpha^6}{360}+\frac{7\alpha^4}{12}+\alpha^2\right)\xi^2\diff{^{2}\Phi^{(1)}}{\xi^{2}}
+ \left(\frac{\alpha^6}{360}+\frac{\alpha^4}{12}+\alpha^2\right)\xi\diff{\Phi^{(1)}}{\xi}
+  6\Phi^{(0)}\Phi^{(1)} = 0,\label{e:Psi1}
\end{align}
Note that this is a linear differential equation, for $\Phi^{(1)}$, and hence the only singular points that appear must already be present in $\Phi^{(0)}$: any singular points in the solution can only be introduced by the nonlinear ODE \eqref{e:Psi0}. The remaining equations can be used to find asymptotic corrections to the singular point locations (see \cite{lustri2023locating}) if required.

\subsubsection{Pad\'{e} Approximation}\label{S.C1.3}

It is difficult to solve \eqref{e:Psi0}. We instead employ Pad\'{e} approximation in order to identify branch points in the solution. The key idea is that we can find a series solution to \eqref{e:Psi0} about $\xi = 0$, and calculate a rational Pad\'{e} approximation for this series. We will then study the distribution of singular points in the Pad\'{e} approximation to infer the form of the singularities in $\Phi^{(0)}$.

We write a series expansion for $\Phi^{(0)}(\xi)$ using \eqref{e:tseries3}, giving
\begin{equation}\label{e:TaylorA}
\Phi^{(0)}(\xi) = \sum_{q=0}^{\infty}\xi^q \Phi_q^{(0)},
\end{equation}
and use the fact that $\Phi_0^{(0)} = 0$ and $\Phi_1^{(0)} = 1$ to obtain the first term in the series. We may then obtain a recurrence relation for the coefficients from \eqref{e:Psi0} by balancing coefficients, giving for $k \geq 2$
\begin{align}\nonumber
\Bigg[\frac{\alpha^6}{360}(k)_6 
+ \frac{\alpha^{6}}{24}(k)_5
+ \left(\frac{13\alpha^6}{72}+\frac{\alpha^4}{12}\right)(k)_4
+ &\left(\frac{\alpha^6}{4}+\frac{\alpha^4}{2}\right)(k)_3\nonumber\\
+ \left(\frac{31\alpha^6}{360}+\frac{7\alpha^4}{12}+\alpha^2\right)(k)_2
&+ \left(\frac{\alpha^6}{360}+\frac{\alpha^4}{12}+\alpha^2\right)k\Bigg] \Phi_k^{(0)}
 =  -3 \sum_{q = 1}^{k-1}\Phi_{q}^{(0)}\Phi_{k-q}^{(0)},\label{e:TaylorRecur}
\end{align}
where $(k)_n$ denotes the $n$th falling factorial of $k$. Note that $\Phi_q^{(0)}$ is not dependent on $z$, being equal to a constant at every order. 

The expression in \eqref{e:TaylorA} gives a Taylor approximation for ${\Phi}^{(0)}$ around $\xi = 0$ (ie. the region in which the exponential contribution in \eqref{e:transpar} is small) whose coefficients are generated by \eqref{e:TaylorRecur}. The terms in \eqref{e:TaylorA} have different exponential scalings in the limit $\eps \to 0$, and therefore are well-ordered and do not interact. This means that the series terms do not contain singularities in the small-$\xi$ region.

To study singularities in ${\Phi}^{(0)}$ in the region where $\xi$ is not small, we use the Taylor expansion to generate a rational function approximation for ${\Phi}^{(0)}$ using Pad\'{e} approximation; see standard references on Pad\'{e} approximation, for example, \cite{graves1981pade}, for the method to compute Pad\'{e} approximation coefficients from Taylor series coefficients. The resultant Pad\'{e} approximant has the form
\begin{equation}\label{e:Pade}
[L/M] = \frac{a_0 + a_1 \xi + \ldots + a_L \xi^L}{b_0 + b_1 \xi + \ldots + b_M \xi^M},
\end{equation}
where the Taylor expansion of $[L/M]$ agrees with the expansion from \eqref{e:TaylorA} up to the coefficient of $\xi^{L+M}$. 

\begin{figure}[tbp]
\centering
\subfloat[Finite-difference family]{
\includegraphics{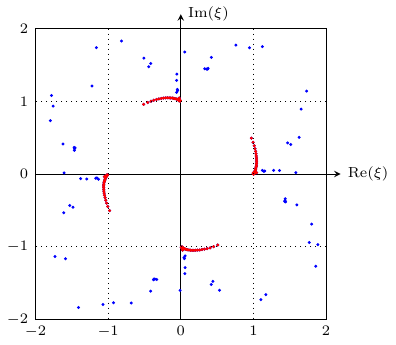}
}
\subfloat[Discrete Painlev\'{e} I family]{
\includegraphics{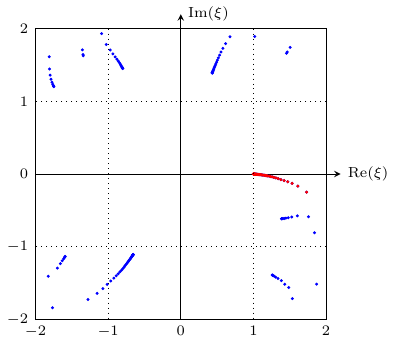}
}
\caption{Poles of the Pad\'{e} approximation with $K=3$ for solutions to the (a) finite-difference family, and (b) discrete Painlev\'{e} I family in terms of the transseries parameter $\xi$. In (a) the red circles denote an accumulation of poles at $\xi = \pm1$ and $\pm\i$, providing numerical evidence for the existence of a branch point at these values of $\xi$. In (b) the red  circles denote an accumulation of poles at $\xi = 1$, providing numerical evidence for the existence of a branch point at this value of $\xi$; {this analysis also predicts additional branch points in the solution that are located outside of the unit circle}. In each figure, the blue circles denote other poles, or accumulations of poles at other values of $\xi$. Note that the $K=1$ case, which corresponds to continuous Painlev\'{e} I, is integrable and does not have branch points in its solutions.}\label{fig:PadeK3}
\end{figure}

These singularities in $\Phi^{(0)}$ can be identified by studying singular points of $[L/M]$, or zeroes of the denominator polynomial; however, we expect the true behaviour of $\Phi^{(0)}$ to contain branch points, and Pad\'{e} approximations in the above form can only ever contain poles. Branch points of $\Phi^{(0)}$ can be identified in Pad\'{e} approximations by locating accumulations of simple poles at the true branch point locations; for more detail on such behaviour, see \cite{stahl1997convergence}.

We evaluated the Taylor expansion \eqref{e:TaylorA} with 1000 decimal places of precision, up to $\Phi_{400}^{(0)}$. These coefficients were used to generate a Pad\'{e} approximant with order polynomials on the numerator and the denominator of order 200. The poles of the Pad\'{e} approximant, found by computing the zeroes of the denominator polynomial, are shown in Figure \ref{fig:PadeK3}(a). This figure shows that the rational approximation contains accumulations of simple poles at $\xi = \pm 1$ and $\pm\i$, providing numerical evidence for the claim that $\Phi^{(0)}$ contains branch points at these locations.

Recall that $\xi$ is defined by \eqref{e:transpar}, and therefore these points correspond to the exponential term being $\mathcal{O}(1)$ in the limit that $\eps \to 0$; hence all of these branch points occur in the red region of Figure \ref{fig:SL3}(b). It is possible to invert this expression to determine the location of these branch points in $z$ (see, for example, \cite{lustri2023locating}); however, this is not necessary for the purposes of this study. Figure \ref{fig:PadeK3} shows numerical evidence for the presence of branch points in the solution $y(z)$, which is expected as the equation \eqref{e:3} is not integrable.

\subsection{Discrete Painlev\'{e} I Family}\label{S:AppDPIK3}

In this section, we outline how the analysis changes if we use the discrete Painlev\'{e} I family \eqref{e:familygen} with $K=3$. This gives the differential equation
\begin{equation}\label{e:DPI3}
\frac{\epsilon^4}{360}\diff{^6 y}{z^6} + \frac{\epsilon^2}{12}\diff{^4 y}{z^4} + \diff{^2 y}{z^2} + 3 y^2  +  \frac{\epsilon^4 y}{12}\diff{^4 y}{z^4}+  \epsilon^2 y \diff{^2 y}{z^2}  =-2z.
\end{equation}
Recall that this equation has the same Stokes structure as the $K=3$ member of the finite-difference family, shown in Figure \ref{fig:SL3}, and the only difference in the exponentially small remainder term \eqref{e:finiteKjump} is the value of $\Lambda$, which changes only the value of $\sigma$ in the transseries expression. 

We use the transseries expression from \eqref{e:tseries}, which we write in terms of the transseries parameter $\xi$ defined in \eqref{e:transpar}. We again select $\Phi_0^{(0)} = 0$, $\Phi_1^{(0)} = 1$, and $\sigma = 2\pi\i\Lambda/\mu$, where $\Lambda$ takes a different value in this problem to the analysis of \eqref{e:3}. The first difference between this and the analysis from Section \ref{S:AppFDK3} is the inclusion of extra terms in the differential equation for $\Phi^{(0)}$. The new equation for $\Phi^{(0)}$ is
\begin{align}
\nonumber
\frac{\alpha^6}{360}\xi^6&\diff{^{6}\Phi^{(0)}}{\xi^{6}}
+ \frac{\alpha^{6}}{24}\xi^5\diff{^{5}\Phi^{(0)}}{\xi^{5}}
+ \left(\frac{13\alpha^6}{72}+\frac{\alpha^4}{12}\right)\xi^4\diff{^{4}\Phi^{(0)}}{\xi^{4}}
+ \left(\frac{\alpha^6}{4}+\frac{\alpha^4}{2}\right)\xi^3\diff{^{3}\Phi^{(0)}}{\xi^{3}}\\
\nonumber
&+ \left(\frac{31\alpha^6}{360}+\frac{7\alpha^4}{12}+\alpha^2\right)\xi^2\diff{^{2}\Phi^{(0)}}{\xi^{2}}
+ \left(\frac{\alpha^6}{360}+\frac{\alpha^4}{12}+\alpha^2\right)\xi\diff{\Phi^{(0)}}{\xi}
+ 3\left(\Phi^{(0)}\right)^2 \\
&+ \frac{\alpha^4}{12}\xi^4\Phi^{(0)}\diff{^{4}\Phi^{(0)}}{\xi^{4}} 
+ \frac{\alpha^{4}}{2}\xi^3\Phi^{(0)}\diff{^{3}\Phi^{(0)}}{\xi^{3}}
+ \left(\frac{7\alpha^4}{12}+\alpha^2\right)\xi^2\Phi^{(0)}\diff{^{2}\Phi^{(0)}}{\xi^{2}}
+ \left(\frac{\alpha^4}{12}+\alpha^2\right)\xi\Phi^{(0)}\diff{\Phi^{(0)}}{\xi^{3}} = 0.
\label{e:Psi0DPI}
\end{align}
We substitute the Taylor expansion \eqref{e:TaylorA} into this expression to obtain a recurrence relation for the coefficients. The new recurrence relation is given for $k \geq 2$ by
\begin{align}\nonumber
&\Bigg[\frac{\alpha^6}{360}(k)_6 
+ \frac{\alpha^{6}}{24}(k)_5
+ \left(\frac{13\alpha^6}{72}+\frac{\alpha^4}{12}\right)(k)_4
+ \left(\frac{\alpha^6}{4}+\frac{\alpha^4}{2}\right)(k)_3\\
 \nonumber
&+ \left(\frac{31\alpha^6}{360}+\frac{7\alpha^4}{12}+\alpha^2\right)(k)_2
+ \left(\frac{\alpha^6}{360}+\frac{\alpha^4}{12}+\alpha^2\right)k\Bigg] \Phi_k^{(0)} \\
 &=  -\Bigg[3 + \frac{\alpha^4}{12}(k)_4 
 + \frac{\alpha^{4}}{2}(k)_3
 + \left(\frac{7\alpha^4}{12} +\alpha^2\right)(k)_2
 + \left(\frac{\alpha^4}{12}+\alpha^2\right)k\Bigg]
 \sum_{q = 1}^{k-1}\Phi_{q}^{(0)}\Phi_{k-q}^{(0)},\label{e:TaylorRecurDPI}
\end{align}

Finally, we use the recurrence relation to generate the Taylor series \eqref{e:TaylorA}, which we in turn use to generate the Pad\'{e} approximant \eqref{e:Pade}. We evaluated the Taylor expansion \eqref{e:TaylorA} with 1000 decimal places of precision, up to $\Phi_{400}^{(0)}$. These coefficients were used to generate a Pad\'{e} approximant with order polynomials on the numerator and the denominator of order 200. The poles of the Pad\'{e} approximant, found by computing the zeroes of the denominator polynomial, are shown in Figure \ref{fig:PadeK3}(b). 

The nearest accumulation of poles appears to be at $\xi = 1$, with several other accumulations further from the origin. This provides numerical evidence for the existence of branch points in $\Phi^{(0)}$ at these locations. As $\xi$ is defined by \eqref{e:transpar},  these points correspond to the exponential term not being small; hence these branch points occur in the red-shaded region of Figure \ref{fig:SL3}(b). 

By comparing Figure \ref{fig:PadeK3}(a) and (b), we see that, while both Pad\'{e} approximants predict that the solution for $\Phi^{(0)}$ contains branch cuts, the inclusion of the extra terms in \eqref{e:DPI3} causes these cuts to change location in the complex-$\xi$ plane. 

\section{Transasymptotic Analysis for General \texorpdfstring{$K$}{K}}\label{s:AppTransGen}

It is relatively straightforward to adapt the analysis from Appendix \ref{s:AppTrans} to study the behaviour of solutions to general members of the finite-difference family \eqref{e:familyFD} and the discrete Painlev\'{e} I family \eqref{e:familygen} for larger values of $K$. The purpose of this analysis is to provide numerical evidence that the red-shaded regions of Figure \ref{fig:stokesK}, and for higher values of $K$, also contains moveable branch points.

While this analysis leads to the equations becoming significantly more complicated, each step can be performed automatically using computational algebra software. Note that $\chi'$ must be recalculated for each value of $K$. In each case, we select the value of $\chi'$ that has the smallest positive real part. When multiple values of $\chi'$ have the same real part (typically $\mathrm{Re}(\chi') = 0$ for even values of $K$), we select the value of $\chi'$ that has the smallest imaginary part. This will be dominant behaviour that appears across the first anti-Stokes line encountered when travelling in a positive angular direction around the singular point, and will therefore be the first exponential contribution to become asymptotically large. These values of $\chi'$ are shown in Table \ref{t:chid}.

\begin{table}
\begin{center}
\begin{tabular}{ |c  c| c  c |} 
 \hline
$K$ & $\chi'$ & $K$ & $\chi'$\\ 
 \hline
 $3$ &  $1.4095506 + 4.1215086\i$  &
$4$ &  $4.6347826\i$  \\ 
 $5$ &  $0.8636477 + 5.2334901 \i$  & 
$6$ &  $5.5426941 \i$  \\ 
$7$ &  $0.4318698 + 6.0148983\i$  &
$8$ &  $6.0889340 \i$  \\ 
$9$ &  $0.0814026 + 6.2726546 \i$  &&\\
 \hline
\end{tabular}
\end{center}
\caption{The value of $\chi'$ corresponding to the first anti-Stokes line encountered when moving in the positive angular direction around a singularity at $z = z_s$. Note that the value is imaginary if $K$ is even, and that the imaginary part of $\chi'$ approaches $2\pi$ as $K$ grows. This is consistent with the behaviour predicted in Figure \ref{fig:largeK}.}
\label{t:chid}
\end{table}

\begin{figure}[tbp]
\centering
\subfloat[$K=4$]{
\includegraphics{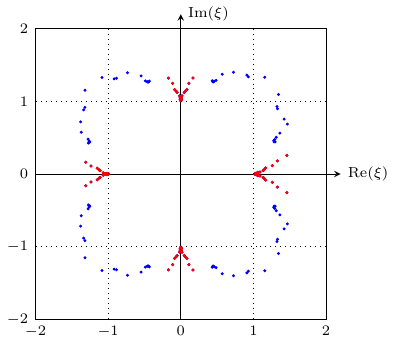}
}
\subfloat[$K=5$]{
\includegraphics{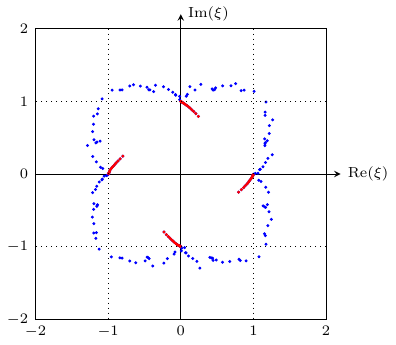}
}
\caption{Poles of the Pad\'{e} approximation for solutions to the finite-difference family with (a) $K=4$ and (b) $K=5$. The red circles denote an accumulation of poles at $\xi = \pm1$ and $\pm\i$, providing numerical evidence for the existence of a branch point at these values of $\xi$. The blue circles denote other poles, or accumulations of poles at other values of $\xi$.}\label{fig:PadeFDgen}
\end{figure}

In this section, we present figures depicting the pole locations for the Pad\'{e} approximation of $\Phi^{(0)}$ for a selection of values of $K>3$, and use this to infer the location of branch points in the complex $\xi$-plane. In each case, we will find a branch point predicted at $\xi = 1$, which corresponds to the behaviour at a location $z$ where the exponential term \eqref{e:transpar} is $\mathcal{O}(1)$ as $\epsilon \to 0$.

Confirming the location of the branch points in $\Phi^{(0)}$ obtained for both the finite-difference and discrete Painlev\'{e} I families would be a worthwhile numerical challenge; however, it is beyond the scope of the present study. The purpose of this analysis was to demonstrate that moveable branch points can emerge in the solution as a consequence of interactions between the growing exponential terms that appear across Stokes lines, confirming that the analysis is consistent with the observation that the equations in each family are not integrable for finite $K$.

\subsection{Finite-Difference Family}
In all cases within this section, we evaluated Taylor expansions with 1000 decimal places of precision, up to $\Phi_{400}^{(0)}$. These coefficients were used to generate Pad\'{e} approximants with polynomials on the numerator and the denominator of order 200, unless otherwise noted. 

We present the Pad\'{e} approximation poles here for $K = 4$ and $5$ for the finite-difference family \eqref{e:familyFD}. For values of $K$ larger than this, it is still possible to identify the branch point locations using the Pad\'{e} approximation method, but it is challenging to depict clearly the location of the accumulation points in $\xi$. The poles of each Pad\'{e} approximant are shown in Figure \ref{fig:PadeFDgen}. For each value of $K$, there are branch points predicted at $\xi = \pm 1$ and $\pm\i$; these are most visibly apparent for $K=4$ and $K=5$, but checking the pole location numerics carefully shows that it is also true for the larger values of $K$. 

Note that these computations suggest the existence of other branch points in the solution; this may be seen most clearly in Figure \ref{fig:PadeFDgen} (a) for $K=4$, where additional pole accumulation points with fewer poles are visible on either side of each of the predicted branch points at $\xi = \pm1$ and $\pm\i$. The pole locations are shown in Figure \ref{fig:PadeFDK4}, where additional pole accumulation points are visible at several points in each quadrant. This plot suggests that $\Phi^{(0)}$ possesses a very complicated singularity structure with numerous branch points. This accumulation of poles in different locations in Figure \ref{fig:PadeFDK4} suggests the possible existence of a natural boundary in the solution, although further analysis would be required in order to prove whether or not such a natural boundary does exist.

\begin{figure}[tbp]
\centering
\includegraphics{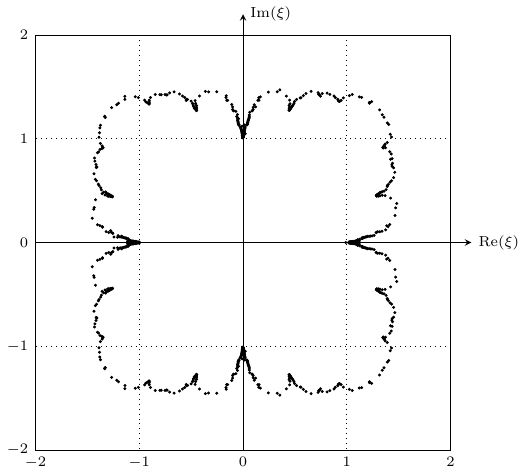}
\caption{Detailed Finite Difference Pad\'{e} for $K=4$. Poles are represented by black circles. In addition to accumulations at $\xi = \pm1$ and $\pm\i$, the figure shows accumulations of poles in numerous locations between these points.}\label{fig:PadeFDK4}
\end{figure}

\subsection{Discrete Painlev\'{e} I Family}

\begin{figure}[tbp]
\centering
\subfloat[$K=4$]{
\includegraphics{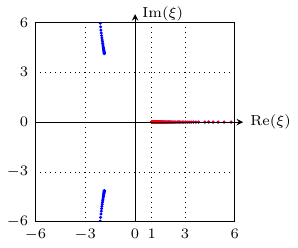}
}
\subfloat[$K=5$]{
\includegraphics{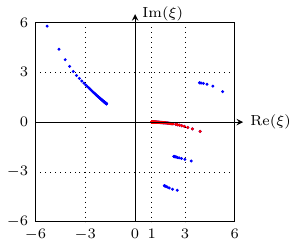}
}
\subfloat[$K=6$]{
\includegraphics{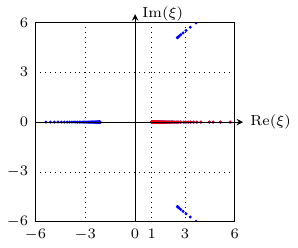}
}

\subfloat[$K=7$]{
\includegraphics{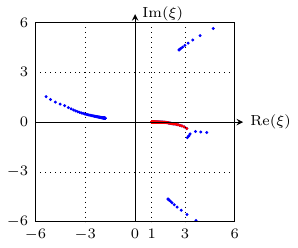}
}
\subfloat[$K=8$]{
\includegraphics{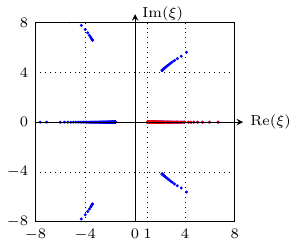}
}
\subfloat[$K=9$]{
\includegraphics{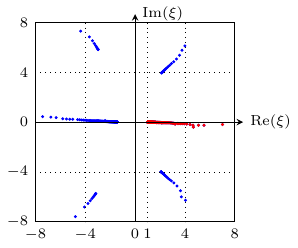}
}
\caption{Poles of the Pad\'{e} approximation for solutions to the discrete Painlev\'{e} I family for values of $K$ ranging from 4 to 9. The red circles denote an accumulation of poles at $\xi = 1$, providing numerical evidence for the existence of a branch point at this value of $\xi$. The blue circles denote accumulations of poles at other values of $\xi$.}\label{fig:PadeDPIgen}
\end{figure}

We present the Pad\'{e} approximation poles here for $K = 4$ to $K=9$ for the discrete Painlev\'{e} family \eqref{e:familygen}. 

For $K = 8$ and $K=9$, we evaluated the Taylor expansion with 2000 decimal places of precision, up to $\Phi_{800}^{(0)}$. These coefficients were used to generate a Pad\'{e} approximant with order polynomials on the numerator and the denominator of order 400. The poles of each Pad\'{e} approximant are shown in Figure \ref{fig:PadeDPIgen}. 

For each value of $K$, there is a branch point predicted at $\xi = 1$, with additional branch points also visible in the solution. The solution possesses vertical symmetry for even values of $K$, and the asymmetry becomes less pronounced for odd values of $K$ as $K$ increases. This appears to be a consequence of the fact that $\chi'$ is imaginary for even $K$, while if $K$ is odd, $\chi'$ has a nonzero real part that decreases in magnitude as $K$ increases. If we instead select $\chi'$ to take the complex conjugate value of the value chosen for Figure \ref{fig:PadeDPIgen}, the resultant figures are reflected vertically in the complex plane.

\section{The inner difference equation for \texorpdfstring{$\Delta = 0$}{Delta = 0} and \texorpdfstring{$\Delta = 1$}{Delta = 1} }\label{S.A2Delta}

Here we consider the homogeneous difference equation
\begin{equation}\label{e:A2_DelEq}
\left(1 + \left(1+\tfrac{\Delta}{2}\right)w_n\right)(w_{n+1} - 2 w_n + w_{n-1}) + 3 w_n^2 = 0,
\end{equation}
where we identify $w_n$ with $v(\eta)$ in the difference equation \eqref{e:discretescaleDel}. We focus on exceptional properties of the special cases $\Delta = 0$ and $\Delta = 1$, with potential implications for their integrability properties. 

To do this, we will identify polynomial conserved quantities that are symemtric properties of $w_n$ and $w_{n+1}$ and contain no higher powers of $w_{n+1}$ than squares. The reasons for these constraints readily become apparent over the course of the calculation. This approach proves effective in precisely two cases, $\Delta = 0$ and $\Delta = 1$, as follows.

\subsection{Case 1: $\Delta = 0$}

The case $\Delta = 0$ is the discrete Painlev\'{e} I equation, and therefore contains known conserved quantities. One of these can be constructed by defining
\begin{equation}\label{e:A2_Fdef0}
F_0(w_n,w_{n+1}) = w_n w_{n+1}(w_n + w_{n+1}) + (w_n - w_{n+1})^2.
\end{equation}
It can be shown that
\begin{equation}
F_0(w_n,w_{n+1})  - F_0(w_{n-1},w_{n}) = (w_{n+1} - w_n)\left[\left(1 + w_n\right)(w_{n+1} - 2 w_n + w_{n-1}) + 3 w_n^2 \right].
\end{equation}
By comparing the right-hand side with \eqref{e:A2_DelEq}, we see that it is equal to zero. Consequently
\begin{equation}\label{e:A2_constF}
F_0(w_n,w_{n+1}) = E
\end{equation}
for constant $E$. 

\subsection{Case 1: $\Delta = 1$}

We can perform a similar analysis for the case $\Delta = 1$. In this case we can derive
\begin{equation}\label{e:A2_Fdef1}
F_1(w_n,w_{n+1}) = -\tfrac{3}{4}w_n^2 w_{n+1}^2 + w_n w_{n+1}(w_n + w_{n+1}) + (w_n - w_{n+1})^2.
\end{equation}
In the same fashion as previously, we calculate
\begin{align}
F_1(w_n,w_{n+1})  - F_1(w_{n-1},w_{n}) &= -\tfrac{1}{2}(w_{n+1} - w_n)(w_n-2)\left[\left(1 + \tfrac{3}{2}w_n\right)(w_{n+1} + w_{n-1}) - 2 w_n \right],\\
&= -\tfrac{1}{2}(w_{n+1} - w_n)(w_n-2)\left[\left(1 +\tfrac{3}{2}w_n\right)(w_{n+1} - 2 w_n + w_{n-1}) + 3 w_n^2\right].
\end{align}
Again, comparing this with \eqref{e:A2_DelEq} reveals that the right-hand side is 0. Consequently
\begin{equation}\label{e:A2_constF1}
F_1(w_n,w_{n+1}) = E
\end{equation}
for constant $E$. 

\subsection{Connecting the two cases}

Returning to the $\Delta = 0$ case, the inner solution we are required to match has $E = 0$. We can determine this by substituting the expression for $v(\eta)$ from \eqref{e:D1series} into an appropriately scaled version of \eqref{e:A2_Fdef0}. Setting $w_n = 1/\sigma_n$ in \eqref{e:A2_constF} implies 
\begin{equation}\label{e:A2_signp1}
\sigma_{n+1} = \tfrac{1}{2}\left(2\sigma_n - 1 \pm (1 - 8 \sigma_n)^{1/2}\right),
\end{equation}
and then introducing $\rho_n = (1-8\sigma_n)^{1/2}$, so that
\begin{equation}\label{e:A2_sig}
\sigma_n = \tfrac{1}{8}(1-\rho_n^2).
\end{equation}
Solving \eqref{e:A2_signp1} after making this substitution gives
\begin{equation}\label{e:A2_rho}
\rho_{n+1} = \pm(\rho_n\pm 2),
\end{equation}
where the signs may be chosen independently. Solving this and substituting the result into \ref{e:A2_sig} gives two possible solutions. If the first sign choice in \eqref{e:A2_rho} is positive, we obtain
\begin{equation}\label{eq:sigma_n_1}
\sigma_n = \tfrac{1}{8} - \tfrac{1}{2}(n - n_0)^2
\end{equation}
for constant $n_0$. Selecting $n_0 = 0$ gives the required inner solution \eqref{e:D0series}. Note that if the first sign choice in \eqref{e:A2_rho} is negative, we instead obtain
\begin{equation}
\sigma_n = A(-1)^n - 2 A^2
\end{equation}
for constant $A$. Note that \eqref{eq:sigma_n_1} is a useful example of a discrete solution that satisfies the singularity confinement property often associated with discrete integrability. Up to translation, the solution is given by
\begin{equation}
w_n = \frac{2}{n(1-n)},
\end{equation}
obtained by setting $n_0 = 1/2$ (being even around $n = 1/2$). In this case, the solution possesses singularities that are confined to a finite region, as they are located at two points ($n = 0$ and $n = 1$).

In the $\Delta = 1$ case where $F_1$ is defined in \eqref{e:A2_Fdef1}, we set $E = 0$ in \eqref{e:A2_constF1} and again define $w_n = 1/\sigma_n$. If we then let $\sigma_n \mapsto \sigma_n + \tfrac{3}{8}$, we recover \eqref{e:A2_signp1} by direct algebraic manipulation. Inverting the map allows us to compare the result with \eqref{e:D1series} and confirm that it is the required inner solution. We can show that this solution possesses the singularity confinement property by setting $n_0 = 1$ (being even around $n = 1$). This reveals that the solution up to translation is given by
\begin{equation}
w_n = \frac{2}{n(2-n)},
\end{equation}
which is a translated version of the solution \eqref{e:A2_signp1}. This solution also possesses singularities located at $n = 0$ and $n = 2$. In this case, the singularities are seperated from each other, but still remain confined to a finite region (this does not of course imply that all solutions have confined singularities; we have not taken the analysis further).

\bibliographystyle{plain}
\bibliography{reference2}

\end{document}